\documentclass[%
 aip,
 amsmath,amssymb,
 reprint,%
nofootinbib]{revtex4-1}

\usepackage{graphicx}
\usepackage{xcolor}
\usepackage{dcolumn}
\usepackage{bm}
\usepackage{cancel}
\usepackage{lineno}

\usepackage[utf8]{inputenc}
\usepackage[T1]{fontenc}
\usepackage{mathptmx}
\usepackage{etoolbox}
\usepackage{comment}
\usepackage[hidelinks]{hyperref}

\newcommand{\Alfven}{Alfv\'{e}n }

\newcommand{\figref}[1]{Fig.~\ref{#1}}
\newcommand{\V}[1]{\mathbf{#1}}

\newcommand{\zhat}{\mbox{$\hat{\mathbf{z}}$}}

\makeatletter
\def\@email#1#2{%
 \endgroup
 \patchcmd{\titleblock@produce}
  {\frontmatter@RRAPformat}
  {\frontmatter@RRAPformat{\produce@RRAP{*#1\href{mailto:#2}{#2}}}\frontmatter@RRAPformat}
  {}{}
}%
\makeatother
\begin{document}

\preprint{AIP/123-QED}

\title[Velocity-Space Signature of Ion Cyclotron Damping]{Unveiling the Velocity-Space Signature of Ion Cyclotron Damping Using Liouville Mapping}

\author{Rui Huang}
\email{rui-huang@uiowa.edu}
 \affiliation{Department of Physics and Astronomy, University of Iowa, Iowa City, IA 52242, USA.}
\author{Gregory G. Howes}%
 \affiliation{Department of Physics and Astronomy, University of Iowa, Iowa City, IA 52242, USA.}

\date{\today}

\begin{abstract}
Ion cyclotron damping is a key mechanism for the dissipation of electromagnetic wave energy in weakly collisional plasmas. This study presents a combined approach using Liouville mapping and the field-particle correlation technique to investigate qualitatively and quantitatively the velocity-space signature of ion cyclotron damping. Liouville mapping offers a computationally efficient way to predict perturbations to the particle velocity distribution function using single-particle trajectories in prescribed electromagnetic fields. One may apply the field-particle correlation technique to these perturbed velocity distributions to reveal the unique velocity-space signatures of the secular energy transfer rate associated with specific wave-particle interactions. We validate this method by reproducing known Landau damping signatures for kinetic Alfvén waves, and then we apply this method to ion cyclotron waves where ion cyclotron damping dominates. The resulting velocity-space signature reveals distinct energization features of ion cyclotron damping : (i) a quadrupolar pattern in the perpendicular $(v_x, v_y)$ plane;  and (ii) a localized energization near the $n = 1$ resonant velocity in gyrotropic $(v_\parallel, v_\perp)$ velocity-space. The quantitative patterns remain unchanged as the ion plasma beta $\beta_i$ is varied, ultimately showing minimal $v_\perp$ dependence on $\beta_i$ of the velocity-space signature at the $n = 1$ resonant velocity. This work provides the first systematic study of how the ion cyclotron damping velocity-space signature varies with $\beta_i$, offering a practical foundation for the future work to identify ion cyclotron damping using kinetic simulation data or spacecraft data. 

\end{abstract}

\maketitle

\section{\label{sec:intro} Introduction}

The dissipation of turbulence and the resulting heating of the plasma species play a fundamental role in the evolution of astrophysical and heliospheric plasmas. Identifying the physical mechanisms underlying the observed steepening of the turbulent energy power spectrum at small scales \cite{kiyani2015dissipation} remains an open question in heliophysics. The physics of this turbulent dissipation involves a number of potential processes\cite{howes2024fundamental}: (1) resonant wave-particle interactions such as Landau damping, transit-time damping, and cyclotron damping; (2) non-resonant wave-particle interactions including stochastic heating, magnetic pumping, and viscous-like damping from kinetic temperature anisotropy instabilities; and (3) dissipation within coherent structures, such as collisionless magnetic reconnection occurring in turbulence-generated current sheets.

Identifying the physical mechanisms in turbulent plasmas is critical for linking the observed spectral features to the underlying physics. Such insight has the potential to enhance our ability to interpret observations of turbulence in space plasmas and is essential to the development of  predictive models of the turbulent plasma heating\cite{howes2024fundamental}. One approach is to compute the energy transfer rate based on the fluid description using measurements of the field and plasma fluctuations and compare that rate to theoretical predictions for specific processes. However, this method often proves inconclusive when multiple turbulent damping mechanisms contribute to the dissipation, especially since those mechanisms generally scale differently with the fundamental parameters of the plasma and turbulence.  Furthermore, this fluid approach underutilizes the rich phase-space information available in the measured velocity distributions of the plasma species.

A more effective approach based on kinetic theory is the field-particle correlation (FPC) technique. First defined by Klein \& Howes (2016) \cite{klein2016measuring}, the FPC has the mathematical form of a modified statistical correlation between the electric field measurements and particle distribution function collected at one spatial point over an observed time interval \cite{howes2017diagnosing}. Physically, the FPC quantifies the velocity-space distribution of the secular energy transfer rate from the electromagnetic field to the particles as a function of the three-dimensional (3D) velocity-space of particle velocity. These velocity-space distributions are found to display characteristic patterns that enable the identification of specific mechanisms, even in cases where multiple mechanisms are acting coincidently \cite{klein2020diagnosing,afshari2024direct}. More importantly, these patterns associated with well-known wave-particle interactions---such as Landau damping or transit-time damping---persist even in turbulent data, enabling a clear identification of the channels of turbulent energy dissipation \citep{klein2017diagnosing,Howes:2018a,TCLi:2019,klein2020diagnosing,Horvath:2020,chen2019evidence,afshari2021importance,conley2023characterizing,huang2024velocity,afshari2024direct}.

To determine whether a specific mechanism contributes to the damping of turbulence in a given simulation or observation, one must first discover that mechanism's unique velocity-space signature. This is typically achieved through well-controlled simulations where that single mechanism dominates by carefully selecting the simulation parameters. The FPC computed from such simulation data reveals a characteristic pattern of the particle energization in velocity space that is then used to identify the mechanism, known as the \emph{velocity-space signature}. Once these velocity-space signatures are established for each proposed damping mechanism, we can analyze the given turbulent datasets. By computing the FPC from the turbulent data and comparing the resulting velocity-space signature with the known signatures, we can determine which physical mechanisms are active. A close match between the turbulent FPC and a known signature provides strong evidence that the corresponding mechanism contributes to energy dissipation in that system.

For instance, the FPC technique has been successfully applied to kinetic numerical simulations to discover the velocity-space signatures of Landau damping \cite{klein2016measuring, howes2017diagnosing, conley2023characterizing}, transit-time damping \cite{huang2024velocity}, cyclotron damping \cite{klein2020diagnosing}, magnetic pumping \cite{montag2022field}, magnetic reconnection \cite{mccubbin2022characterizing}, and collisionless shocks \cite{juno2021field,juno2023phase,Brown:2023,Howes:2025}. Guided by these revealed signatures, the FPC technique has been utilized to identify these physical mechanisms in real-world data, from lab experiments \cite{schroeder2021laboratory} to spacecraft observations \cite{chen2019evidence, afshari2024direct,Montag:2025}.

From a methodological perspective, the prerequisite for applying the FPC technique to identify physical mechanisms from turbulent datasets is the discovery of unique velocity-space signatures corresponding to each mechanism.
This initial ``signature discovery'' step is often limited by the cost and complexity of generating well-controlled data, particularly the velocity distribution function data. For example, Klein \emph{et al.} (2020) \cite{klein2020diagnosing} discovers the velocity-space signature of ion cyclotron damping using hybrid Vlasov-Maxwell simulations of Alfvénic-ion turbulence, but these simulations are computationally expensive and offer limited flexibility to explore parameter dependencies.

A promising way to address this constraint is the Liouville mapping technique, as detailed in Schwartz \emph{et al.} (1998)\cite{schwartz19987}. Based on Liouville's theorem, which states that the phase-space distribution function remains constant along a particle's trajectory, this method reconstructs the distribution function by tracking single-particle orbits in known electromagnetic fields, bypassing the direct solving of the Vlasov-Maxwell equation system and enabling each run to be performed on a standard laptop rather than supercomputers. This computational efficiency allows systematic parameter scans and rapid numerical testing, tasks that would be impractical using fully self-consistent kinetic codes. Previous work \cite{schroeder2021laboratory} applies this method to calculate FPC for one-dimensional Landau damping in comparison with laboratory measurements. To date, however, the Liouville mapping method remains underexplored as a general approach to predict the velocity-space signatures of different turbulent damping mechanisms, particularly in three-dimensional contexts.

From a physics perspective, resonant wave-particle interactions are widely recognized as fundamental processes for turbulent energy transfer \cite{chen2019evidence, afshari2021importance, afshari2024direct}, yet important gaps remain. For example, the 3D velocity-space signature of ion cyclotron damping and its dependence on the ion plasma beta $\beta_i$ (the ratio of the ion thermal pressure to the magnetic pressure) has not been systematically explored. We focus on $\beta_i$ because it is not only the most important parameter across all turbulent damping mechanisms but also the specific key parameter determining whether ion cyclotron damping occurs and where the associated energization patterns appear in velocity space \cite{howes2024fundamental}.

Motivated by these considerations, this study employs the Liouville mapping technique to generate well-controlled data for exploring the detailed velocity-space signatures of two resonant wave-particle interactions: Landau damping and cyclotron damping. The Landau damping case, having been thoroughly studied in previous work, serves primarily to validate the method, while the cyclotron damping case is the focus of new physics uncovered by our method. This paper is structured as follows. Section~\ref{sec:fpc} provides a detailed overview of diagnosing the secular energy transfer from the electromagnetic field to the particles in resonance using FPC. Section~\ref{sec:lm} describes how the Liouville mapping technique is incorporated into the FPC computation. Section~\ref{sec:ld_results} presents results for ion Landau damping as the method validation. Section~\ref{sec:cd_results} displays the key findings of this work: the detailed velocity-space signatures of ion cyclotron damping. The properties of the used ion cyclotron wave (ICW) modes are outlined in Section~\ref{sec:icw_properties}. The velocity-space signature presented in the $(v_x, v_y)$ and $(v_\parallel, v_\perp)$ planes is shown in Section~\ref{sec:cd_1w_2w}, with corresponding interpretations given in Section~\ref{sec:cd_cexcey_interpret} and Section~\ref{sec:cd_ceperp_interpret}. Finally, Section~\ref{sec:cd_beta} presents the first analysis of the dependence of ion cyclotron damping velocity-space signature on $\beta_i$.

The definitions of important physical quantities used in this paper can be found in Table~\ref{tab:params}. 

\begin{table*}
\caption{\label{tab:params} Definitions of Important Physical Quantities Used in this Paper}
\begin{ruledtabular}
\begin{tabular}{clcl}
Quantity & Definition & Quantity & Definition\\
\hline
$s$ & Particle species & $v_\text{ts} = \sqrt{2 T_{0s} / m_s}$ & Particle thermal speed   \\
$m_s$ & Mass & $c$ & Speed of light \\
$q_s$ & Charge & $\Omega_s = q_s B_0 / (m_s c)$ & Species cyclotron frequency\\
$\mathbf{v}$ & Particle velocity & $\rho_s = v_\text{ts}/\Omega_s$ & Species Larmor radius \\
$\mathbf{B}_0$ & Equilibrium magnetic field & $v_A = B_0 / \sqrt{4 \pi m_i n_{0i}}$ & \Alfven speed \\
$n_{0s}$ & Species number density & $\beta_i = 8 \pi n_{0 i} T_{0 i} / B_0^2 = v_\text{ti}^2 / v_A^2$ & Ion plasma beta \\
$T_{0s}$ & Species temperature (units: eV) & $\mathbf{k} = k_\perp \hat{x} + k_\parallel \hat{z}$ & Wave vector\\
$f_s(\mathbf{r},\mathbf{v}, t)$ & Distribution function & $\omega$, $\gamma$& Wave frequency and growth rate\\
$\mathbf{E}(\mathbf{r}, t)$ & Electric field & $w_s (\mathbf{r},\mathbf{v}, t) \equiv (1/2) m_s v^2 f_s$ & Phase-space energy density\\
$\mathbf{B}(\mathbf{r}, t)$ & Magnetic field & $W_s (\mathbf{r}, t) \equiv \int w_s d \mathbf{v}$ & Energy density\\
$\mathbf{J}_s(\mathbf{r}, t)$ & Species current density & $\mathcal{W}_s (t) \equiv \int W_s d \mathbf{r}$ & Energy\\
\end{tabular}
\end{ruledtabular}
\end{table*}
\section{\label{sec:fpc} The Physics of Resonant Wave-Particle Interactions}
The primary objective of this study is to characterize resonant wave-particle interactions using single-point measurement data. These interactions between the charged particles and plasma waves satisfy the resonance condition
\begin{eqnarray}
\omega - k_\parallel v_\text{res, n} = n \Omega_s.
\label{eq:res_condition}
\end{eqnarray}
The left-hand side of Eq.~(\ref{eq:res_condition}) represents the Doppler-shifted wave frequency in the reference frame moving with the particle's parallel velocity; the resonance occurs when that Doppler shifted frequency is equal to a harmonic of the cyclotron frequency $\Omega_s$ of particle species $s$, where the integer $n = 0, \pm 1, \pm 2, ...$ arises from a Bessel expansion for the plane wave \citep{stix1992waves}.
Here $\omega$ is the wave frequency and $\V{k}$ is wavevector, where $k_\parallel$ is its component along the equilibrium magnetic field direction, $\V{B}_0=B_0 \zhat$.  Note that the $n = 0$ resonance corresponds to Landau resonance, the $|n| > 0$ resonance correspond to the fundamental and harmonics of cyclotron resonance. Within the scope of this study, which focuses on parallel-propagating wave modes, only $n = 0, \pm 1$ resonances are expected \cite{verscharen2013dispersion}. 

To analyze these interactions, we employ the field-particle correlation technique, which is rooted in the Vlasov equation describing the evolution of the distribution function $f_s(\mathbf{r}, \mathbf{v}, t)$ for particles of species $s$ in a collisionless plasma:
\begin{eqnarray}
\frac{\partial f_s}{\partial t} + \mathbf{v} \cdot \frac{\partial f_s}{\partial \mathbf{r}} + \frac{q_s}{m_s} \left(\mathbf{E} + \frac{\mathbf{v} \times \mathbf{B}}{c} \right) \cdot \frac{\partial f_s}{\partial \mathbf{v}} = 0
\label{eq:vlasov_eq}
\end{eqnarray}
The interaction between the electromagnetic fields and the particle distribution function is encoded in the third term on the left-hand side of the equation.
Multiplying by $(1/2) m_s v^2$ and defining the phase-space energy density as $w_s (\mathbf{r},\mathbf{v}, t) \equiv (1/2) m_s v^2 f_s$, we obtain its evolution:
\begin{eqnarray}
    \frac{\partial w_s}{\partial t} = - \mathbf{v} \cdot \frac{\partial w_s}{\partial \mathbf{r}} - \frac{1}{2} q_s v^2 \mathbf{E} \cdot \frac{\partial f_s}{\partial \mathbf{v}} - \frac{1}{2} q_s v^2  \frac{\mathbf{v} \times \mathbf{B}}{c} \cdot \frac{\partial f_s}{\partial \mathbf{v}}
\label{eq:ws_eq}
\end{eqnarray}
Within a chosen spatial volume, the total microscopic kinetic energy contained in the particles, $\mathcal{W}_s = \int w_s d \mathbf{v} d \mathbf{r}$, changes only due to the electric field in the second term. Integrating the electric field term over velocity space yields the rate of change of energy density of species $s$ due to work done by the electric field, $\mathbf{J}_s \cdot \mathbf{E}$, where $\mathbf{J}_s$ denotes the current density from the motion of particles of species $s$.  The advection term (first term) vanishes under either infinite or periodic spatial boundary conditions, and the magnetic term (third term) redistributes particles in velocity space but does not alter the total energy \citep{howes2017diagnosing}.

Based on the electric field term in Eq.~(\ref{eq:ws_eq}), the field-particle correlation (FPC) technique is developed. The correlations for species $s$, $C_{E_x, s}, C_{E_y, s}, C_{E_\parallel, s}$ and $C_{E_\perp, s}$, computed at a fixed spatial location $\mathbf{r}_0$ and centered at time $t_0$, are defined as
\begin{eqnarray}
&& C_{E_x, s} (\mathbf{r}_0, \mathbf{v}, t_0; \tau) \nonumber \\
&& \equiv \frac{1}{\tau} \int_{t_0 - \tau/2}^{t_0 + \tau/2} \left[ - q_s \frac{v_x^2}{2} \frac{\partial f_s(\mathbf{r}_0, \mathbf{v}, t)}{\partial v_x} E_x (\mathbf{r}_0, t) \right]  \ d t, 
\label{eq:cex}
\end{eqnarray}
\begin{eqnarray}
&& C_{E_y, s} (\mathbf{r}_0, \mathbf{v}, t_0; \tau) \nonumber \\
&& \equiv \frac{1}{\tau} \int_{t_0 - \tau/2}^{t_0 + \tau/2} \left[ - q_s \frac{v_y^2}{2} \frac{\partial f_s(\mathbf{r}_0, \mathbf{v}, t)}{\partial v_y} E_y (\mathbf{r}_0, t) \right] \ d t, 
\label{eq:cey}
\end{eqnarray}
\begin{eqnarray}
&& C_{E_\parallel, s} (\mathbf{r}_0, \mathbf{v}, t_0; \tau) = C_{E_z, s} (\mathbf{r}_0, \mathbf{v}, t_0; \tau)  \nonumber \\
&& \equiv \frac{1}{\tau} \int_{t_0 - \tau/2}^{t_0 + \tau/2} \left[ - q_s \frac{v_z^2}{2} \frac{\partial f_s(\mathbf{r}_0, \mathbf{v}, t)}{\partial v_z} E_z (\mathbf{r}_0, t) \right] \ d t, 
\label{eq:cepar}
\end{eqnarray}
\begin{eqnarray}
&& C_{E_\perp, s} (\mathbf{r}_0, \mathbf{v}, t_0; \tau) \nonumber \\
&& \equiv C_{E_x, s} (\mathbf{r}_0, \mathbf{v}, t_0; \tau) + C_{E_y, s} (\mathbf{r}_0, \mathbf{v}, t_0; \tau) 
\label{eq:ceperp}
\end{eqnarray}
where these correlations are time-averaged over an appropriately chosen correlation interval $\tau$. Within the scope of this work, we consider a uniform background magnetic field $\mathbf{B}_0$ aligned with the $z$-axis, which we also refer to as the parallel direction. Accordingly, the subscripts $\parallel$ and $z$ are used interchangeably: $\parallel$ appears when paired with $\perp$, while $z$ appears when paired with $x$ and $y$. Note that the $v^2$ in the electric field term in Eq.~(\ref{eq:ws_eq}) is replaced in these definitions by $v_j^2$, where the field component is indicated by the index $j$---this replacement yields no change in the velocity-integrated rate of change of phase-space energy density \citep{klein2017diagnosing}.

Each of the correlations in Eqs.~(\ref{eq:cex})--(\ref{eq:ceperp}) quantifies how the different components of the electric field ($E_x$, $E_y$, $E_\parallel$ or $E_\perp$) contribute to the time-averaged rate of change of the phase-space energy density $w_s(\mathbf{r}_0, \mathbf{v}, t_0)$. With a suitably chosen correlation interval $\tau$, the field-particle correlation isolates the secular energy transfer between the electric field and the particles by the cancellation of any oscillatory energy transfer over its characteristic oscillation period \citep{howes2017diagnosing}.  
When integrated over 3D velocity space, the FPC yields a quantitative determination of the rate of change of spatial energy density, $W_s(\V{r_0},t) = \int w_s(\V{r_0},\V{v},t) d \mathbf{v} $.  

At a particular spatial point $\mathbf{r}_0$ and time $t_0$, the field-particle correlation $C_{E_j, s}(\mathbf{r}_0, \mathbf{v}, t_0; \tau)$ due to electric field component $E_j$ produces a pattern representing the rate of energy transfer to the particles throughout 3D velocity space, known as the \emph{velocity-space signature} of the particle energization.  
The FPCs in 3D velocity space are usually expressed in either cylindrical $(v_\parallel, \theta, v_\perp)$ or Cartesian $(v_x, v_y, v_z)$ coordinates. For simplicity, we use the notation $C_{E_j, s} (v_\parallel, \theta, v_\perp)$ or $C_{E_j, s}(v_x, v_y, v_z)$ with $j = x, y, \parallel, \perp$, and explicitly denote $\mathbf{r}_0$, $t_0$, and $\tau$ only when necessary.

The velocity-space signatures generated by the FPC technique are most easily visualized on a two-dimensional (2D) plane that emphasizes the features of the particular kinetic energization mechanism. For example, previous studies show that plotting $C_{E_\parallel, s}$ on the gyrotropic plane $(v_\parallel, v_\perp)$, denoted as $C_{E_\parallel, s}(v_\parallel, v_\perp)$, captures the features of Landau damping \citep{klein2016measuring, howes2017diagnosing, conley2023characterizing}, while plotting $C_{E_x, s}$, $C_{E_y, s}$, and $C_{E_\perp, s}$ on the Cartesian velocity plane $(v_x, v_y)$, denoted as $C_{E_j, s}(v_x, v_y)$ where $j = x, y, \perp$, reveals helpful characterizations associated with cyclotron damping \citep{afshari2024direct}.

Ideally, the distribution function is obtained as a dataset in the full 3D velocity space, yielding the calculated FPC over the same domain. To visualize these results on a 2D velocity plane, one integrates over the less relevant velocity coordinate. For the examples above, the full cylindrical 3D correlation is integrated over the gyroangle $\theta$ to obtain $C_{E_\parallel, s}(v_\parallel, v_\perp)$ for Landau damping; and the full Cartesian 3D correlation is integrated over the parallel dimension to obtain $C_{E_j, s}(v_x, v_y)$ for cyclotron damping. We refer this approach as \emph{including the full 3D velocity space information}, since the 2D patterns are derived by integrating a complete 3D dataset.
However, in some cases, practical constraints---such as limited computational resources or simplified modeling assumption---may restrict the distribution function to a 2D domain in which one velocity axis is neglected, yielding FPC patterns that are inherently 2D and thus can be plotted directly. We refer to this as \emph{neglecting the dependence on the ``omitted” velocity axis}.

Generally, different kinetic mechanisms of particle energization yield unique velocity-space signatures that can be used to identify the mechanism from single-point, time-series measurements of the 3D velocity distributions $f_s(\mathbf{r}_0, \mathbf{v}, t)$ and the electric field $\mathbf{E}(\mathbf{r}_0, t)$.  Below, we describe how we use Liouville mapping to obtain these quantities in computationally efficient way.

\section{\label{sec:lm} Liouville Mapping Procedure}

\subsection{\label{sec:lm_overview} Overview}
In plasma physics, the Liouville theorem in Hamiltonian mechanics takes the form of the Vlasov equation, Eq.~(\ref{eq:vlasov_eq}), where the left hand side of represents the Lagrangian derivative along a particle trajectory in phase space, while the right hand side is zero.
This implies that the distribution function remains constant along a particle's trajectory in six-dimensional phase space. Specifically,
\begin{eqnarray}
f_s(\mathbf{r}(t_1), \mathbf{v} (t_1), t_1) = f_s(\mathbf{r}(t_2), \mathbf{v} (t_2), t_2),
\end{eqnarray}
where $t_1$ and $t_2$ are two different time samples, and the position $\mathbf{r}(t)$ together with the velocity $\mathbf{v}(t)$ gives the particle's trajectory in phase space.
To track these trajectories, we numerically integrate the following single particle motion equations
\begin{eqnarray}
    \frac{d \mathbf{r}}{dt} = \mathbf{v},
    \label{eq:dxdt}
\end{eqnarray}
\begin{eqnarray}
    \frac{d \mathbf{v}}{dt} = \frac{q_s}{m_s} \left(\mathbf{E} + \frac{\mathbf{v}}{c} \times \mathbf{B} \right).
    \label{eq:dvdt}
\end{eqnarray}
This allows us to map the distribution function between different times if the spatial and temporal variations of the electromagnetic fields are known \cite{schwartz19987}.
While conceptually straightforward, its practical implementation for FPC calculation involves several technical steps, as outlined below.
\subsection{\label{sec:lm_EM}Constructing electromagnetic fields}
The electromagnetic fields for Liouville mapping are constructed by inverse Fourier transforming numerically obtained eigenfunctions. 
These eigenfunctions are computed as functions of the wavevector $\mathbf{k} = k_\perp \hat{x} + k_\parallel \hat{z}$ using the FORTRAN 90 code \texttt{PLUME} \cite{klein2015predicted, klein2025plume}, which solves the dispersion relation of the linearized Vlasov-Maxwell equations in Fourier space, describing a collisionless, hot, uniformly magnetized ion-electron plasma, as formulated in Chapter 10, Eqs. (66)–(73) of Stix (1992) \citep{stix1992waves}.
\texttt{PLUME} requires specification of five key dimensionless input parameters: the normalized parallel and the perpendicular components of the wavevector, $k_\parallel \rho_i$ and $k_\perp \rho_i$; ion plasma beta $\beta_i$; ion-to-electron temperature ratio $T_i/T_e$; and normalized ion thermal velocity $v_{ti}/c$. We specify protons as the ion species by choosing the ion-to-electron mass ratio $m_i/m_e=1836$.
Outputs from \texttt{PLUME} include the wave frequency $\omega$ and growth or damping rate $\gamma$ (positive for growth, negative for damping) normalized to the ion cyclotron frequency $\Omega_i$, as well as the complex Fourier coefficients of the vector electric and magnetic field eigenfunctions.

To present how we construct the electromagnetic fields in space and time, we let $\mathbf{F}$ denote either the electric field $\mathbf{E}$ or magnetic field $\mathbf{B}$, and express each total field as a sum of an equilibrium component and a perturbation,
\begin{eqnarray}
\mathbf{F} (\mathbf{r}, t) = \mathbf{F}_0 + W(t; t_\text{init}, t') \sum_{l = 1}^n \delta \mathbf{F}_l (\mathbf{r}, t).
\label{eq:constructed_em_fields}
\end{eqnarray}
Here the first term, $\mathbf{F}_0$, is the equilibrium field, with $\mathbf{E}_0 = 0$ and $\mathbf{B}_0 = B_0 \hat z$. The second term represents the total perturbed field due to a superposition of $n$ linear wave modes, where $\delta \mathbf{F}_l (\mathbf{r}, t)$ is the perturbed field contributed by the $l$-th wave mode, given by

\begin{eqnarray}
\delta \mathbf{F}_l (\mathbf{r}, t) = \frac{\epsilon_l}{2} \left \{ \delta \hat{\mathbf{F}} (\mathbf{k}_l) \exp \left[i \left(\mathbf{k}_l \cdot \mathbf{r} - \omega(\mathbf{k}_l) t + \alpha_l \right) \right] \right . \nonumber \\
\left . + \delta \hat{\mathbf{F}}^* (\mathbf{k}_l) \exp \left[-i \left(\mathbf{k}_l \cdot \mathbf{r} - \omega(\mathbf{k}_l) t + \alpha_l \right) \right] \right\},
\label{eq:em_ift}
\end{eqnarray}
where $\mathbf{k}_l$, $\omega(\mathbf{k}_l)$ and $\delta \hat{\mathbf{F}}(\mathbf{k}_l)$ are the wavevector, real wave frequency, and complex field Fourier coefficients for the $l$-th mode, respectively. The hat symbol $\hat{\ }$ indicates quantities in Fourier space, the notational dependence $(\mathbf{k}_l)$ emphasizes that these quantities are functions of the wavevector, and the asterisk $^*$ denotes the complex conjugate. An arbitrary real phase $0 \le \alpha_l < 2 \pi$ is allowed for each wave mode $l$ to adjust the relative phase among multiple modes. Note that we neglect the growth or damping rate $\gamma(\mathbf{k}_l)$ for all modes so that the each wave mode retains a constant peak amplitude in time.  The amplitude for each wave mode is directly controlled by a dimensionless amplitude factor $\epsilon_l$, defined by
\begin{eqnarray}
\epsilon_l = \frac{c \delta \hat{E}_x(\mathbf{k}_l)}{v_{ti} B_0},
\end{eqnarray} 
where $v_{ti} = \sqrt{2 T_{0i}/m_i}$ is the ion thermal speed. In the magnetohydrodynamic (MHD) limit, this amplitude factor $\epsilon_l$ is related to the magnitude of the magnetic field perturbation of an \Alfven wave normalized to the equilibrium field by $\delta \hat{B}_y / B_0 = \pm c \sqrt{\beta_i} \delta \hat{E}_x / (v_{ti} B_0)$. This relation can be derived from the eigensolution of the linearized ideal MHD equations, assuming the wavevector lies in the $(\hat{x}, \hat{z})$ plane \cite{klein2012using}. The Alfvén mode satisfies $\delta \hat{B}_y / B_0 = \pm \delta \hat{U}_y / v_A$, where the $E \times B$ drift provides the dominant contribution to the flow velocity, $\delta \hat{U}_y = -c\delta \hat{E}_x / B_0$.

The time-dependent window function $W(t; t_\text{init}, t')$ smoothly ramps up the perturbation over duration $t'$
\begin{eqnarray}
\text{W}(t; t_\text{init}, t') = \begin{cases}
\sin^2 \left(\frac{\pi t}{2 t'} \right), & t_\text{init} \leq t \leq t_\text{init} + t'\\
1, & t > t_\text{init} + t',
\end{cases}
\label{eq:win}
\end{eqnarray}
Here, $t_\text{init}$ marks the start of system evolution.
This gradual ramp-up of the wave amplitude eliminates artifacts in the perturbed distribution function associated with the Liouville mapping. By mapping back to a time $t_\text{init}$ at which the wave amplitude is zero, one may map to a chosen analytical form for the equilibrium velocity distribution function \emph{in the absence of waves}, meaning that the initial distribution is uniform in space.  Since particles starting at the same position but with different velocities will generally be mapped to different positions in space over the same time interval, this would lead to undesired perturbations in the velocity distribution function unless the wave amplitude is zero at  $t_\text{init}$.

In this study, we consider two types of field configurations: (i) a single propagating wave mode with wavevector $\mathbf{k}$, and (ii) a pair of counter-propagating waves with wavevectors $\mathbf{k} = (k_x, k_y, \pm k_z)$, forming a standing wave along the $z$-direction.
In both cases, the perturbed field is periodic with $T = 2 \pi / \omega$. We set $t' = T$ in the window function, Eq.~(\ref{eq:win}), so that the field amplitude ramps up over its first wave period.

This procedure enables the analytical specification of the full spatial and temporal variation of the electric and magnetic fields, $\mathbf{E}(\mathbf{r}, t)$ and $\mathbf{B}(\mathbf{r}, t)$, where the form of Eq.~(\ref{eq:constructed_em_fields}) effectively leads to a linear superposition of plane waves that pervade all space.  This specification also yields 
 the first key quantity needed for computing FPC, the electric field $\mathbf{E}(\mathbf{r}_0, t)$. The second key quantity, the distribution function, will be obtained via Liouville mapping, as detailed in Subsection~\ref{sec:lm_f}.
\subsection{\label{sec:lm_f}Liouville Mapping the Distribution Function}

To reconstruct $f_s(\mathbf{r}_0, \mathbf{v}, t)$ on a pre-determined velocity space grid suitable for FPC computation, we integrate the single particle motion equations, Eqs.~(\ref{eq:dxdt}) and~(\ref{eq:dvdt}), backward in time, starting from the desired time, $t_f$. This approach allows us to trace each particle trajectory from the 
final state $(\mathbf{r}_0, \mathbf{v}, t_f)$ back to its initial state $(\mathbf{r}_\text{init}, \mathbf{v}_\text{init}, t_\text{init})$, when the initial equilibrium velocity distribution function is known.
Here we explicitly emphasize that, since we are integrating the particle trajectories backwards in time, the numerical integration begins at $t_f$ and is followed backwards in time to $t_\text{init}$.

Specifically, we will compute the distribution function at time $t_f$ based on a known initial equilibrium distribution function $f_{s, \text{init}}(\mathbf{r}_\text{init}, \mathbf{v}_\text{init},t_\text{init})$. The window function ensures that the electromagnetic field perturbations are zero at $t_\text{init}$, so that this initial equilibrium distribution function is uniform in space.
In this study, $f_{s, \text{init}}$ is set to be a Maxwellian and thus depends only on velocity.  Recall that, to eliminate the oscillatory energy transfer associated with undamped wave motion, the FPC is calculated by a time-average over a chosen correlation interval $\tau$. Therefore, to evaluate FPCs at a spatial location $\mathbf{r}_0$ and physical time $t_0$, the time interval over which we need to determine the distribution function must span the range $t_0 - \tau/2 < t < t_0 + \tau/2$.  We require knowledge of the distribution function at the \emph{same position} $\mathbf{r}_0$ across this time interval on a pre-defined 3D velocity-space grid with $M$ discrete points, $\{\mathbf{v}_{f1}, \mathbf{v}_{f2}, \dots, \mathbf{v}_{fm}, \dots, \mathbf{v}_{fM}\}$.

We divide the time interval into $N$ equal-length subintervals, representing each with its starting time. This yields $N$ sampled time slices $\{t_{f1}, t_{f 2}, \dots, t_{fn}, \dots,  t_{fN}\}$, with timestep (i.e., the subinterval length) $\Delta t = t_{f n + 1} - t_{f n} = \tau/N$. Under this convention, the first time slice is $t_{f1} = t_0 - \tau/2$ and the last is $t_{fN} = t_0 + \tau/2 - \Delta t$.
Here, the subscript "f" denotes physical final times for each subinterval.

\begin{figure}
    \begin{center}
       \includegraphics*[width=1.0\linewidth]{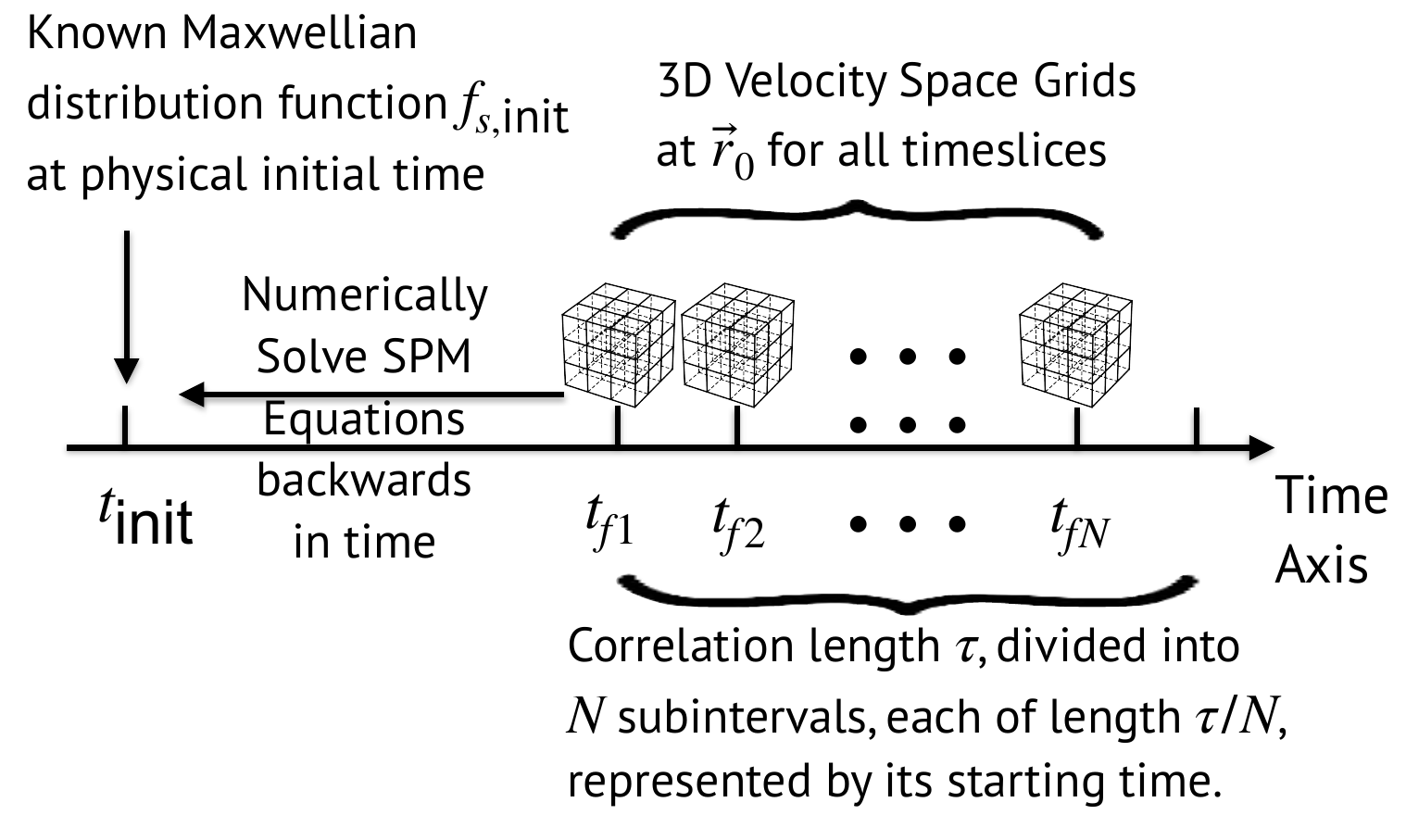} 
    \end{center}
    \caption{Illustration of the procedure for Liouville mapping of the velocity distribution function by integrating the single-particle motion (SPM) equations backwards in time from final state $(\mathbf{r}_0, \mathbf{v}, t_f)$  back to initial state  $(\mathbf{r}_\text{init}, \mathbf{v}_\text{init}, t_\text{init})$. This procedure must be completed for each of the $M$ points in the 3D velocity-space grid, and for each of the $N$ subintervals that span the correlation interval $\tau$.}
    \label{fig:f_construct}
\end{figure}

At the first time slice $t = t_{f1}$, for each velocity grid point $\mathbf{v}_{fm}$, we initialize the final state as $(\mathbf{r} = \mathbf{r}_0, \mathbf{v} = \mathbf{v}_{fm})$ and numerically trace the single particle trajectory backward to the initial time $t_\text{init}$. This yields $(\mathbf{r}_\text{init, m}, \mathbf{v}_\text{init, m})$, from which we use Liouville's theorem to obtain
\begin{eqnarray}
f_s(\mathbf{r}_0, \mathbf{v}_{fm}, t_{f1}) = f_{s, \text{init}}(\mathbf{r}_\text{init, m}, \mathbf{v}_\text{init, m}, t_\text{init}).
\end{eqnarray}
Iterating this process over all $M$ velocity-space points gives the distribution function over the 3D velocity space at $t = t_{f1}$, $f_s(\mathbf{r}_0, \mathbf{v}, t_{f1})$.
Next, we advance to $t_{f2}$, and repeat the above procedure to get $f_s(\mathbf{r}_0, \mathbf{v}, t_{f2})$.
The procedure is then repeated at each subsequent time slice $t_{fn}$, ultimately reconstructing the distribution function over the full time interval, $f_s(\mathbf{r}_0, \mathbf{v}, t_0 - \tau/2 < t < t_0 + \tau/2)$.  This procedure for constructing $f_s$ over the full correlation interval $\tau$ is illustrated in \figref{fig:f_construct}. Thus, we have obtained the second key quantity, the particle velocity distribution function, needed to compute the FPC, as described in Section~\ref{sec:fpc}.
\subsection{\label{sec:lm_fpc}Computing FPC}
With both the electric field and distribution function available from the previous steps, FPCs can be computed. For example, to evaluate $C_{E_\parallel, s}(\mathbf{r}_0, \mathbf{v}, t; \tau)$ as defined in Eq.~(\ref{eq:cepar}), we proceed as follows. 
At each time slice $t_{fn}$, the velocity derivative $\partial f_s / \partial v_z$ is computed numerically using a second-order central finite difference scheme for interior points and first-order forward/backward differences at the boundaries of the velocity grid. This derivative is then multiplied by the factor $- q_s v_z^2 / 2$ and the corresponding electric field component $E_z(\mathbf{r}_0, t_{fn})$ to yield the instantaneous contribution to the FPC, i.e. the integrand in Eq.~(\ref{eq:cepar}). Repeating this calculation across all $N$ time slices within the correlation interval produces a sequence of such contributions, which are then multiplied with the timestep $\Delta t$, summed and divided by the interval length $\tau$ to obtain the final time-averaged FPC. This process can be similarly applied to other components of the electric field, yielding correlations such as $C_{E_x, s}$, $C_{E_y, s}$, or $C_{E_\perp, s}$, depending on the physics under investigation.

\begin{figure*}
    \begin{center}
    \includegraphics[width=1.0\linewidth]{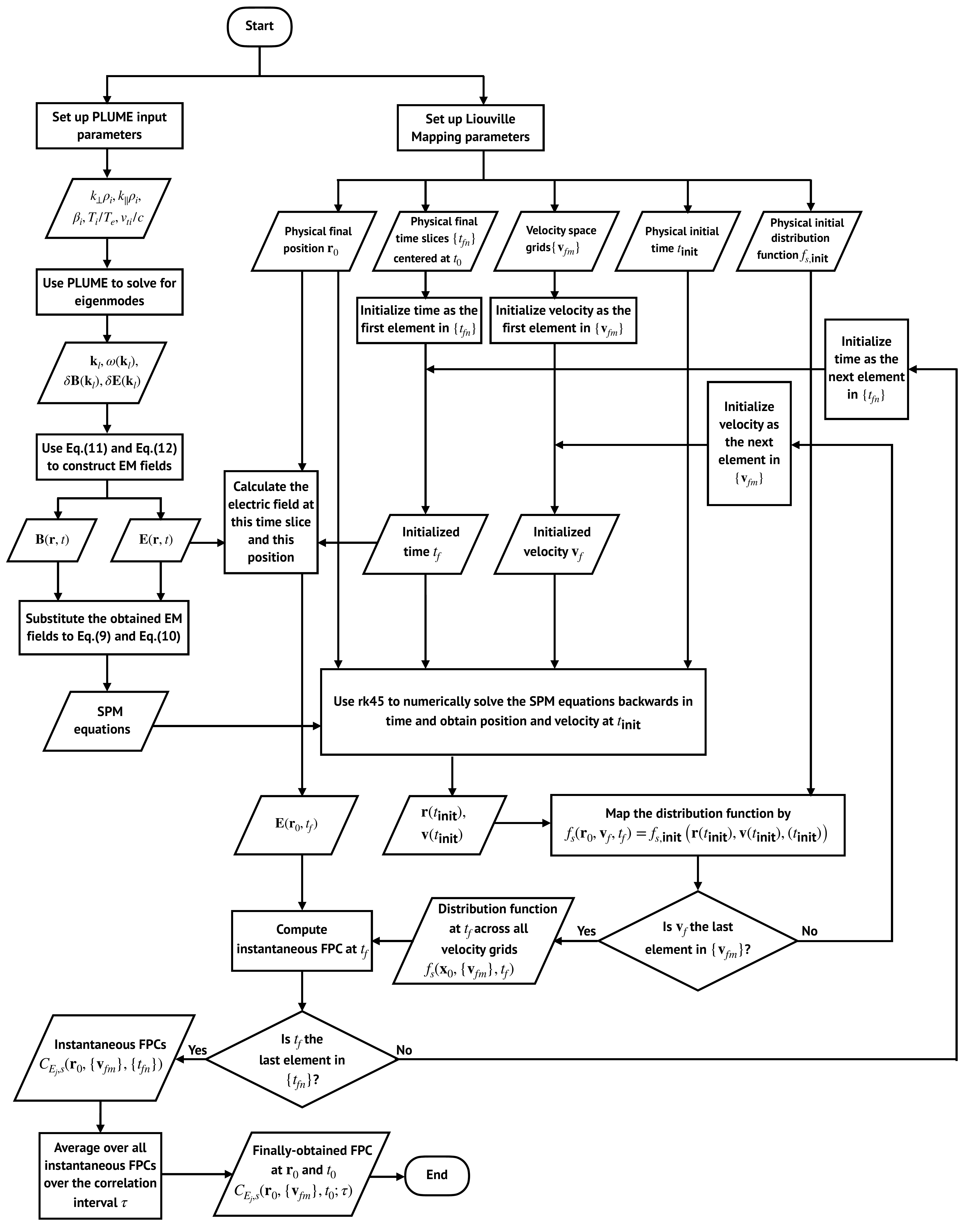}
    \end{center}
    \caption{Flowchart describing the full procedure to construct the electromagnetic fields, perform Liouville mapping of the distribution function, and computation of the FPC, as described in Sec.~\ref{sec:lm}. Here, ``instantaneous FPCs" refer to the integrands in the definitions of FPCs, i.e. Eq.~(\ref{eq:cex})-(\ref{eq:ceperp}), and ``rk45" denotes the Runge-Kutta-Fehlberg method.}
    \label{fig:lmfpc_workflow}
\end{figure*}

In \figref{fig:lmfpc_workflow}, we present a flowchart describing the entire procedure, including the field construction, Liouville mapping of the distribution function, and evaluation of the FPC.
\subsection{\label{sec:lm_misc}Caveats}

First, in a self-consistent plasma a linear wave damps at the damping rate returned by \texttt{PLUME}, yet our calculation keeps its peak amplitude fixed. This might seem artificial, but in turbulent plasmas, many modes are fed by energy from larger scales, so holding one mode at constant strength is a reasonable assumption.
Second, because the electromagnetic wave is pre-defined, we ignore how the particle motion may change the fields. For the small wave amplitude $\epsilon_l$ used here, the perturbation to the particle distribution remains weak, so the distribution function's zeroth- and first-order components that set the linear wave frequency and the phase velocity are essentially unchanged. Variations in the fields from the particle motion appear primarily in the damping rate, not in the wave frequency, leaving the imposed fields qualitatively acceptable to leading order.
Third, the approach fully retains nonlinear dynamics: Eqs.~(\ref{eq:dxdt})–(\ref{eq:dvdt}) are integrated without approximation, and the Liouville mapping keeps all nonlinear terms in the Vlasov equation. This feature becomes crucial when multiple modes are present. As shown in Appendix~\ref{sec:icd_asymmetry}, the FPC signature calculated from the two counter-propagating waves scenario is not a simple superposition of the single-mode results.
Fourth, the magnitudes of the velocity-space signatures in this work are presented in arbitrary units. We are primarily interested in using this technique to predict the qualitative patterns of energization in velocity space and in quantifying the position of those features in velocity space but not the amplitude of those signatures; to make a quantitative comparison of the amplitude of the FPC, one needs to convert our dimensionless quantities into appropriately scaled dimensional quantities.


\section{\label{sec:ld_results}Results: Landau damping}
The velocity-space signature of Landau damping, identified using the FPC technique, has been well established in previous studies \cite{klein2016measuring, howes2017diagnosing, klein2017diagnosing, conley2023characterizing}. A key characteristic of this signature, shown in Figure 6(c) of Klein, Howes, \& TenBarge (2017)\citep{klein2017diagnosing}, is a bipolar pattern centered at the resonant velocity.
To validate our method, we compute the FPC from Liouville mapping data, using the electromagnetic fields set according to the eigenfunction for a kinetic \Alfven wave (KAW) obtained from \texttt{PLUME}, and then we seek the known velocity-space signature of Landau damping.

\subsection{\label{sec:kaw_properties}Properties of Kinetic \Alfven Waves}
\begin{figure}
    \centering    \includegraphics[width=1.\linewidth]{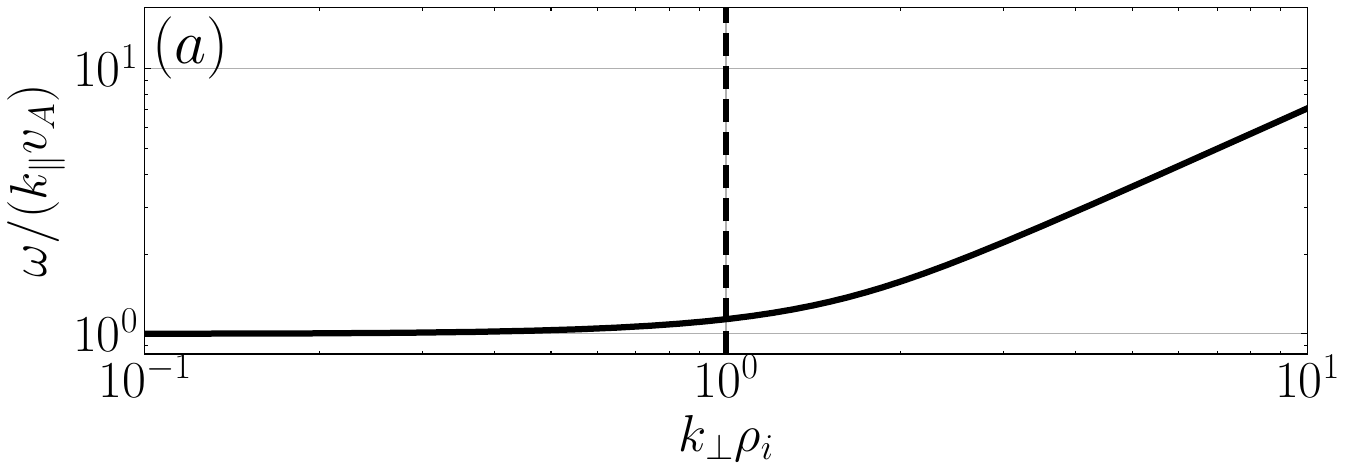}
    \includegraphics[width=1.\linewidth]{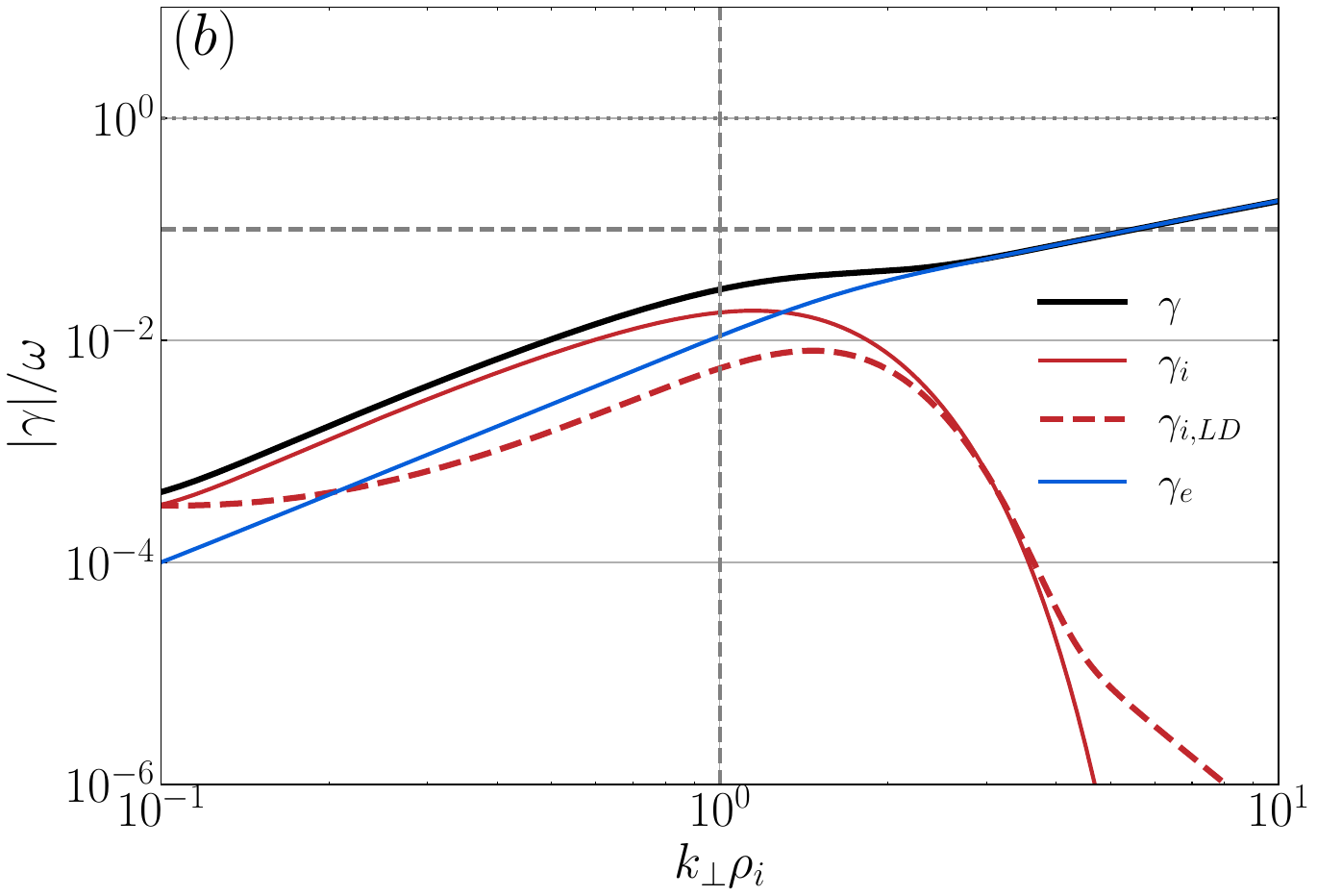}
    \caption{ The normalized (a) frequency $\omega / (k_\parallel v_A)$ and (b)  damping rates $|\gamma|/\omega$ for a KAW in plasma with parameters $\beta_i = 1$, $T_i/T_e = 1$, $m_i/m_e = 1836$, $v_\text{ti} / c = 10^{-4}$.  The KAW wave mode with $k_\parallel \rho_i = 0.05$ is plotted over the range $10^{-1} \leq k_\perp \rho_i  \leq 10^1$. The damping rates plotted include the total damping rate  $|\gamma|/\omega$ (black), ion damping rate $|\gamma_i|/\omega$ (red solid), electron damping rate $|\gamma_e|/\omega$ (blue), and damping rate due to only ion Landau damping $|\gamma_{i,LD}|/\omega$ (red dashed).}
    \label{fig:kaw_mode}
\end{figure}

We plot the linear dispersion relation for the KAW obtained from \texttt{PLUME} for a proton-electron plasma with isotropic equilibrium temperatures for each species and parameters $\beta_i = 1$, $T_i/T_e = 1$, $m_i/m_e = 1836$, $v_\text{ti} / c = 10^{-4}$, and $k_\parallel \rho_i = 0.05$ over the range $10^{-1} \leq k_\perp \rho_i  \leq 10^1$ in \figref{fig:kaw_mode}.  
We plot (a)  the wave frequency normalized by the MHD Alfvén wave frequency $\omega / (k_\parallel v_A)$ and (b) the normalized total damping rate $|\gamma|/\omega$ (black), ion damping rate $|\gamma_i|/\omega$ (red solid), electron damping rate $|\gamma_e|/\omega$ (blue), and separated damping rate due to ion Landau damping $|\gamma_{i,LD}|/\omega$ (red dashed) as calculated by \texttt{PLUME} \citep{huang2024velocity,klein2025plume}, each as a function of $k_\perp \rho_i$. 
The normalized frequency $\omega / (k_\parallel v_A) \simeq 1$ in the MHD limit ($k_\perp \rho_i \ll 1$), and the electric field polarization\citep{verscharen2013dispersion} $\mathcal{P}_E \simeq 0$ indicates linear polarization over the full range plotted, as expected for the KAW.

To determine the electromagnetic field eigenfunctions for a KAW, we choose $k_\perp \rho_i = 1$ near the peak in the ion Landau damping rate from \texttt{PLUME} results, as marked by the vertical dashed lines in both panels of \figref{fig:kaw_mode}. Specifying the wavevector $(k_x \rho_i, k_y \rho_i, k_z \rho_i) = (1, 0, 0.05)$, we obtain the Fourier coefficients of the eigenfunctions $\delta \V{\hat{E}}(\V{k})$ and $\delta \V{\hat{B}}(\V{k})$ and the wave frequency $\omega / \Omega_i = 0.05675$, leading to a period of $T \Omega_i = 110.72$ and a wave phase velocity $\omega / (k_\parallel v_{ti}) = 1.135$. For this mode, we choose phase $\alpha = 0$ and set the amplitude scaling factor to $\epsilon = 0.02$.  The time-averaged Poynting vector, $\V{S}=(c/4\pi) \V{E}\times\V{B}$, yields an electromagnetic energy flux for this KAW that is dominantly along the equilibrium magnetic field direction ($+ \hat{z}$), with just 3\% of the energy flux perpendicular to the magnetic field, so we denote this as the \emph{forward-propagating wave}.

By only changing the sign of the $k_z$ component of the wavevector, giving $(k_x \rho_i, k_y \rho_i, k_z \rho_i) = (1, 0, -0.05)$, we obtain a \emph{backward-propagating wave} with the similar properties
except that the parallel-phase velocity flips sign to $\omega / (k_\parallel v_{ti}) = -1.135$ and the 
electromagnetic energy flux for this KAW is primarily along $- \hat{z}$.   In our Liouville mapping calculations, in addition to the forward-propagating and backward-propagating KAW cases, we also obtain a \emph{counterpropagating KAW case} by linear superposing both of these solutions, yielding a standing KAW for our case of equal counterpropagating wave amplitudes.

\subsection{\label{sec:ld_validate} Landau damping Signatures with Single Kinetic \Alfven Wave and Two Counter-Propagating Kinetic \Alfven Waves}
Next, we compute the gyrotropic parallel FPC, $C_{E_\parallel}(v_\parallel,v_\perp)$ \footnote{From this point forward, we omit the species subscript ``s'' for both the FPCs and the distribution function, as we focus exclusively on ions.}, for the three cases above: (i) forward-propagating KAW, (ii) backward-propagating KAW, and (iii) counterpropagating KAW cases.
For the forward-propagating single-KAW field, we perform Liouville mapping with an  initial time of $t_\text{init} = - 3 T$, and calculate the FPC centered at the final time $t_0 = T/2$ over a correlation interval of one wave period, $\tau = T$. The correlation interval is sampled with $N=40$ time slices, giving a subinterval size of $\Delta t = 0.025T$. The FPC is computed at the single-point $\mathbf{r}_0/\rho_i = (0.1, 0.1, 0.1)$, but for a correlation interval that samples the full $2 \pi$ phase of a traveling wave, the resulting FPC is independent of the measurement position.
 
Because previous studies have shown the velocity-space signature of ion Landau damping varies primarily in the parallel direction of velocity space, we use Liouville mapping to compute the ion distribution function on a velocity-space grid optimized to extract the expected features. Using cylindrical velocity coordinates $(v_\perp, \theta, v_z)$, we sample the parallel velocity with higher resolution, using 64 equally spaced points over $-3.048 \leq v_z/v_{ti} \leq 3.048$; a lower resolution of 10 points is used to sample the perpendicular coordinate over $0 \leq v_\perp/v_{ti} \leq 3$. The physics of Landau damping is independent of the gyrophase (since it is fully described within gyrokinetic theory, which is integrated over gyrophase \citep{Howes:2006}), so we simply choose a single azimuthal slice at $\theta = 0$.  

\begin{figure*}
 \begin{center}
    \includegraphics[width=.33\textwidth]{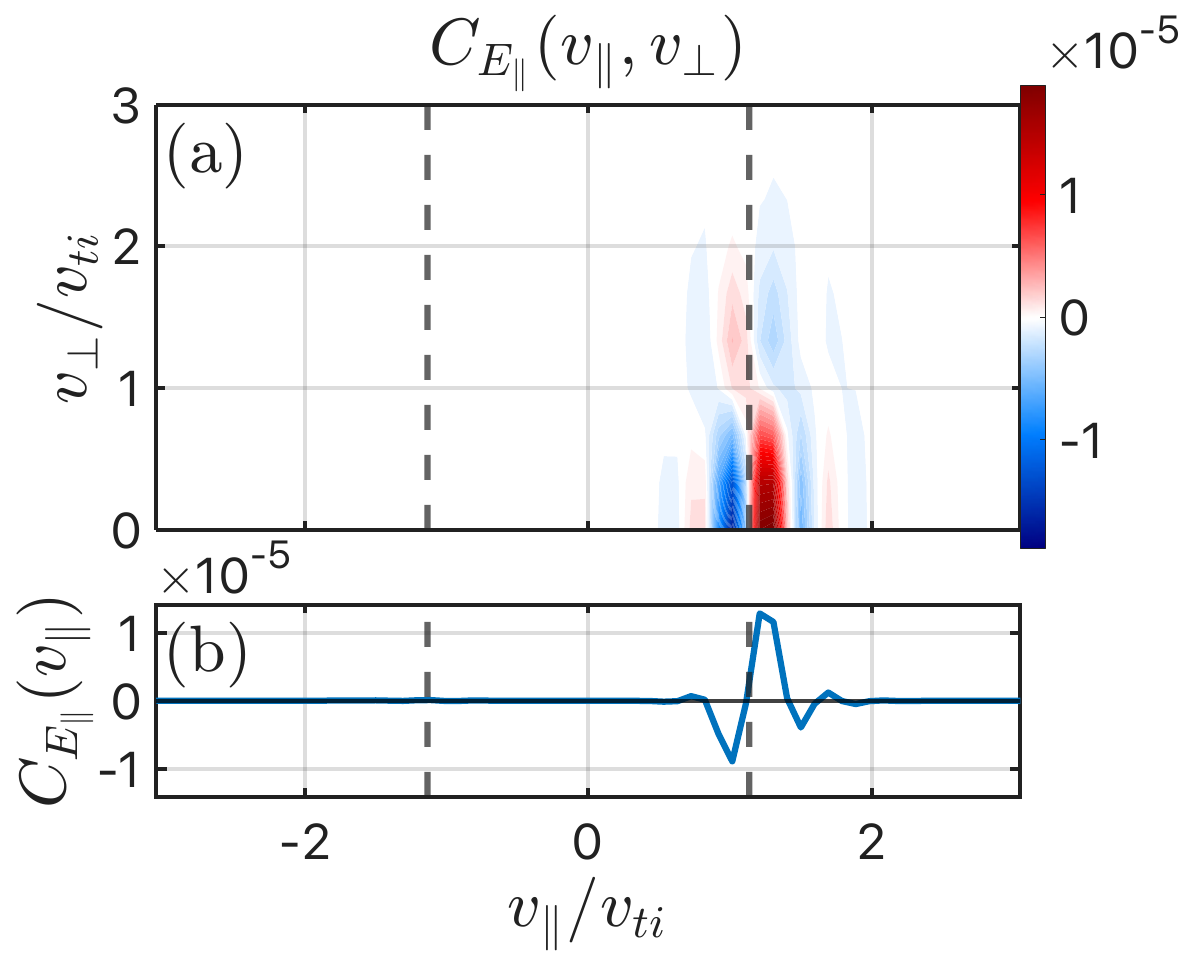}
    \includegraphics[width=.33\textwidth]{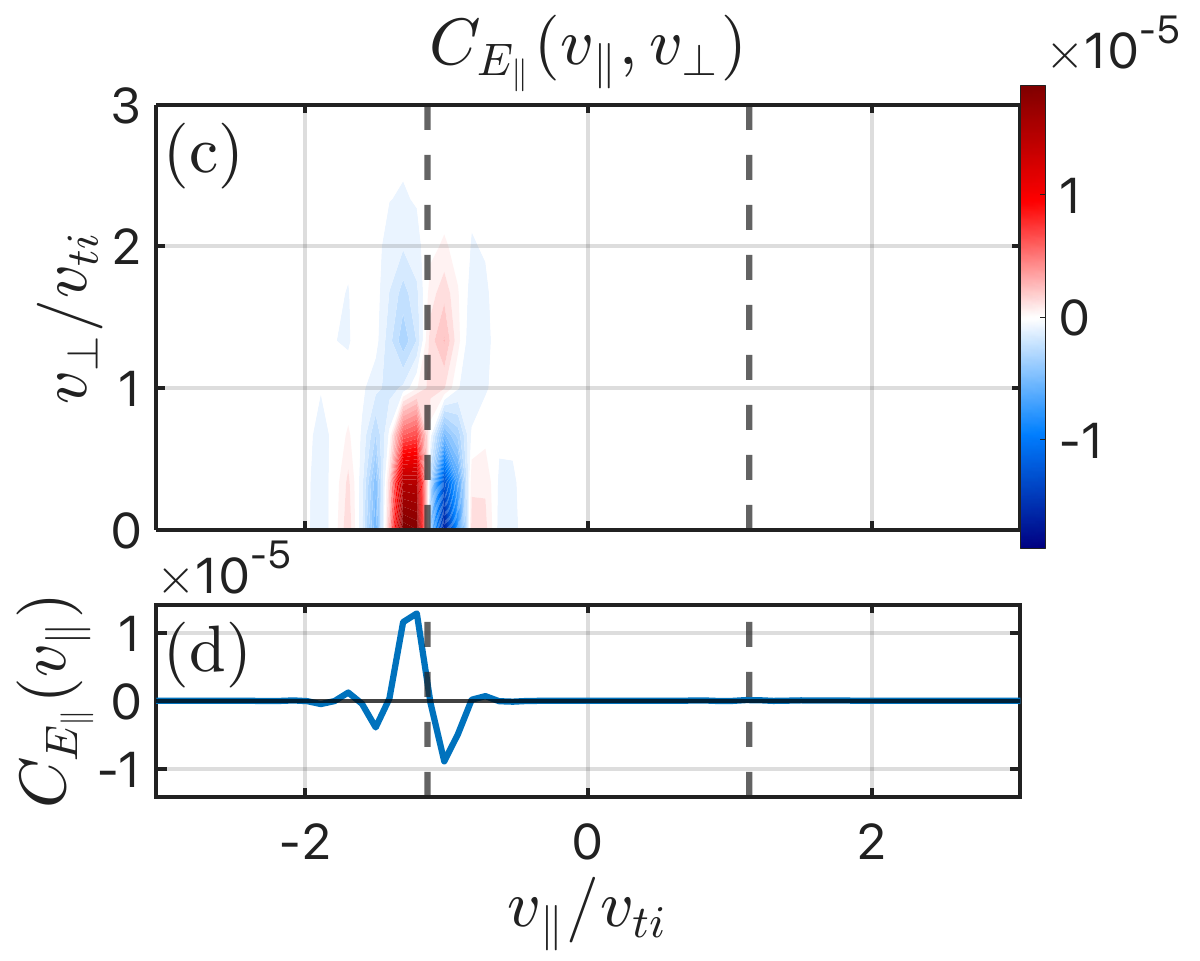}
    \includegraphics[width=.33\textwidth]{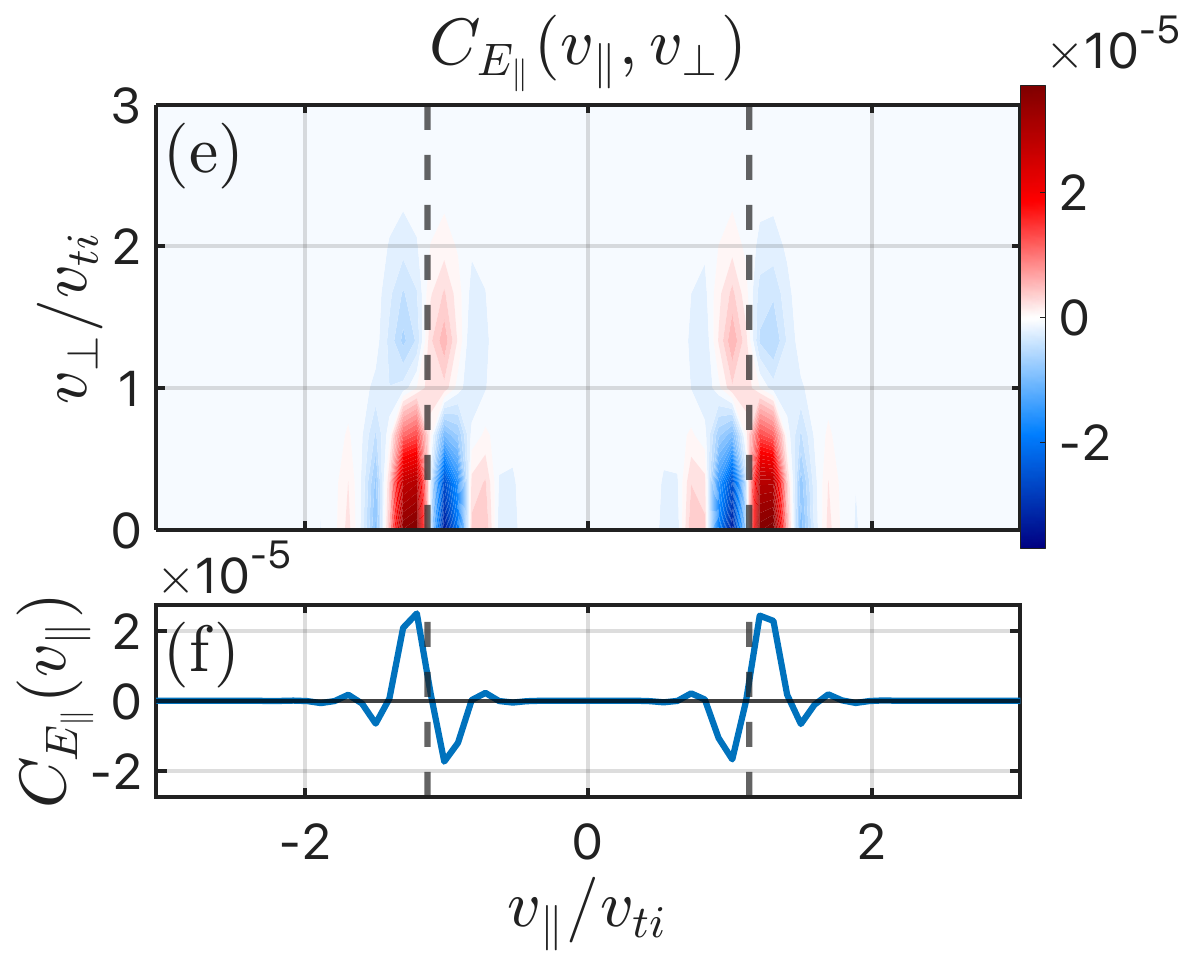}
   \end{center}
\caption{\label{fig:ld_1kaw_2kaw} The gyrotropic parallel FPC $C_{E_\parallel}(v_\parallel,v_\perp)$ (top panels) and reduced parallel FPC $C_{E_\parallel}(v_\parallel)$ (bottom panels) for (a,b) the forward-propagating KAW case, (c,d) the backward-propagating KAW case, and (e,f) the counterpropagating KAW case, showing bipolar velocity-space signatures arise in the directions corresponding to the wave propagation direction, as expected theoretically.}
\end{figure*}

The gyrotropic parallel FPC $C_{E_\parallel}(v_\parallel,v_\perp)$ is plotted in \figref{fig:ld_1kaw_2kaw}(a), recovering the previously discovered velocity-space signature of ion Landau damping \citep{klein2017diagnosing,Howes:2017c,Howes:2018a,klein2020diagnosing}, providing a validation for our Liouville mapping technique.  The bipolar velocity-space signature shows a loss of phase-space energy density (blue) lower than the parallel phase velocity of the wave $\omega/(k_\parallel v_{ti})=1.135$ (vertical dashed line) and a gain of phase-space energy density (red) above the parallel phase velocity, consistent with the expected flattening of the velocity distribution function about the phase velocity due to ion Landau damping \citep{klein2017diagnosing,Howes:2017c}.  With more particles gaining energy than losing it, the net effect is a transfer of energy from the wave to the particles. This bipolar signature is easily observed in the reduced parallel correlation, $C_{E_\parallel}(v_\parallel) = \int C_{E_\parallel}(v_\parallel, v_\perp) d v_\perp$, plotted in \figref{fig:ld_1kaw_2kaw}(b), where the zero-crossing of the signature is coincident with the parallel phase velocity. Note that for this forward-propagating wave, the resonant velocity lies on the positive side of the velocity space, so the entire signature appears on the right.

For the same Liouville mapping parameters, the backward-propagating KAW case shows a bipolar signature appears at $v_\parallel <0$ as expected, showing the gyrotropic parallel FPC $C_{E_\parallel}(v_\parallel,v_\perp)$ in \figref{fig:ld_1kaw_2kaw}(c) and the reduced parallel FPC $C_{E_\parallel}(v_\parallel)$ in \figref{fig:ld_1kaw_2kaw}(d).

For the counterpropagating KAW case with equal wave amplitudes, the resulting electromagnetic fluctuations yield a standing wave pattern. Since it is the parallel electric field component $E_\parallel$ that mediates the energy transfer in Landau damping \cite{schroeder2021laboratory}, we compute the FPC at a position $z/\rho_i = 31.0$, chosen to coincide with an anti-node of the standing $E_\parallel$ fluctuation.
Keeping all other Liouville mapping parameters unchanged, we successfully recover the Landau damping signature at $\mathbf{r}_0/\rho_i = (0.1, 0.1, 31.0)$, as shown in \figref{fig:ld_1kaw_2kaw}(e) and~(f). In this case, we obtain a bipolar signature at both positive and negative parallel velocities, indicating that ions traveling near the phase velocity in both directions resonantly gain energy from the wave when averaged over the correlation interval $\tau$.

Together, the results for the forward-propagating and backward-propagating KAWs and the counterpropagating KAW constitute a thorough validation of our Liouville mapping technique to predict the velocity-space signatures of collisionless damping by the Landau resonance.

\section{\label{sec:cd_results} Results: Cyclotron Damping}
The velocity-space signature of cyclotron damping on the $(v_\parallel, v_\perp)$ plane was first identified by Klein \emph{et al.} (2020) \cite{klein2020diagnosing} from turbulence simulation data.
Subsequently, Afshari \emph{et al.} (2024) \cite{afshari2024direct} characterized the features of ion cyclotron damping on the perpendicular $(v_x, v_y)$ plane by modeling the evolution of the distribution function as a Maxwellian distribution centered at the ion bulk flow velocity in the $(v_x, v_y)$ plane.
However, Klein \emph{et al.} (2020) \cite{klein2020diagnosing} focused solely on turbulent scenarios and did not show how the cyclotron damping signature appears for a single wave mode. While turbulence is more relevant to real-world plasma environments, examining the physics of a single wave mode offers valuable insight into the fundamental physics of ion cyclotron damping.
Furthermore, the model in Afshari \emph{et al.} (2024) \cite{afshari2024direct} is overly simplified and neglects $v_z$ dependence of the velocity-space signature entirely.
More importantly, the variation of these ion cyclotron damping signatures with different $\beta_i$ values has not yet been explored, providing an ideal first science application of the Liouville mapping technique introduced here.

In this section, we address these gaps using our computationally efficient and highly controllable Liouville mapping method.
We begin by presenting the properties of selected ICW modes from \texttt{PLUME} eigenfunction solutions in Section~\ref{sec:icw_properties}.
In Section~\ref{sec:cd_1w_2w}, we reveal the velocity-space signatures of ion cyclotron damping for two single ICW modes with different propagation directions as well as the two counter-propagating ICW modes. 
The following Sections~\ref{sec:cd_cexcey_interpret} and \ref{sec:cd_ceperp_interpret} then explain the physical meaning of these signatures.
Finally, in Section~\ref{sec:cd_beta}, we examine how these signatures vary with $\beta_i$.

\subsection{\label{sec:icw_properties} Properties of Ion Cyclotron Wave modes}
\begin{figure}
    \centering
    \includegraphics[width=.8\linewidth]{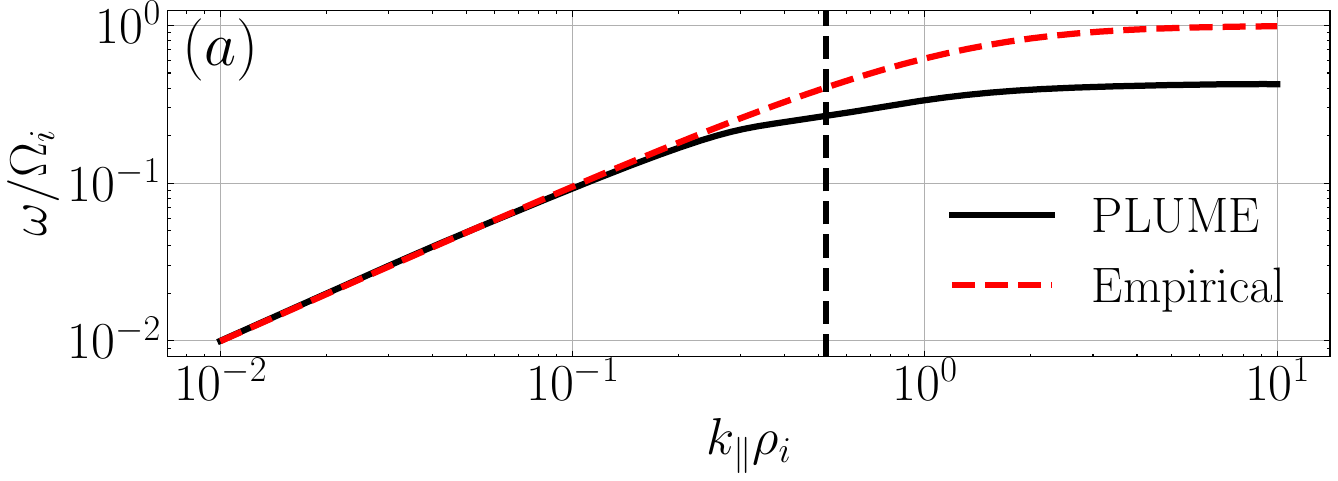}
    \includegraphics[width=.8\linewidth]{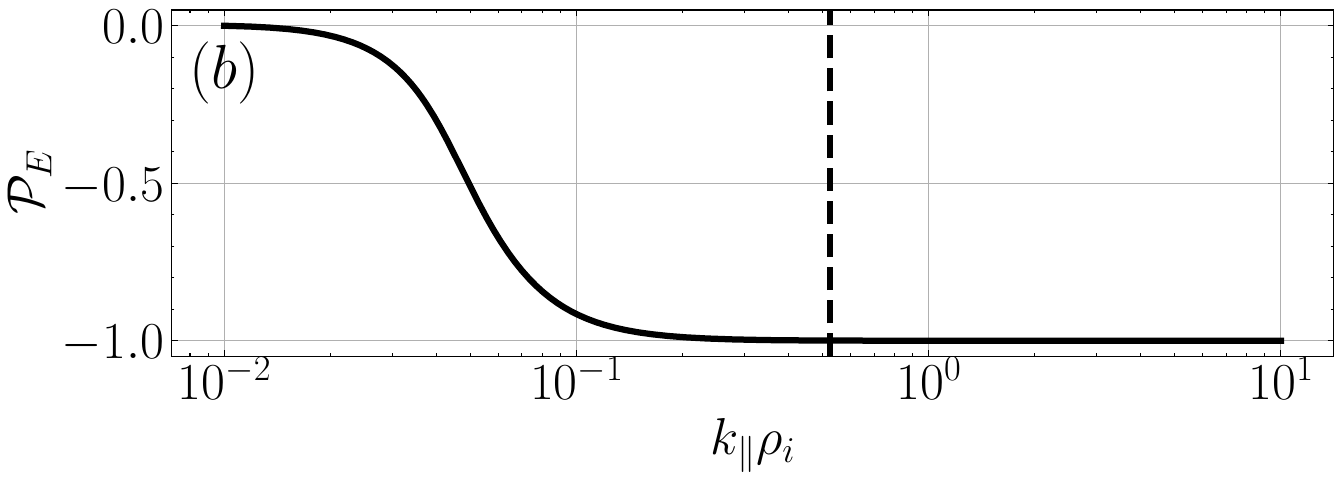}
    \includegraphics[width=.8\linewidth]{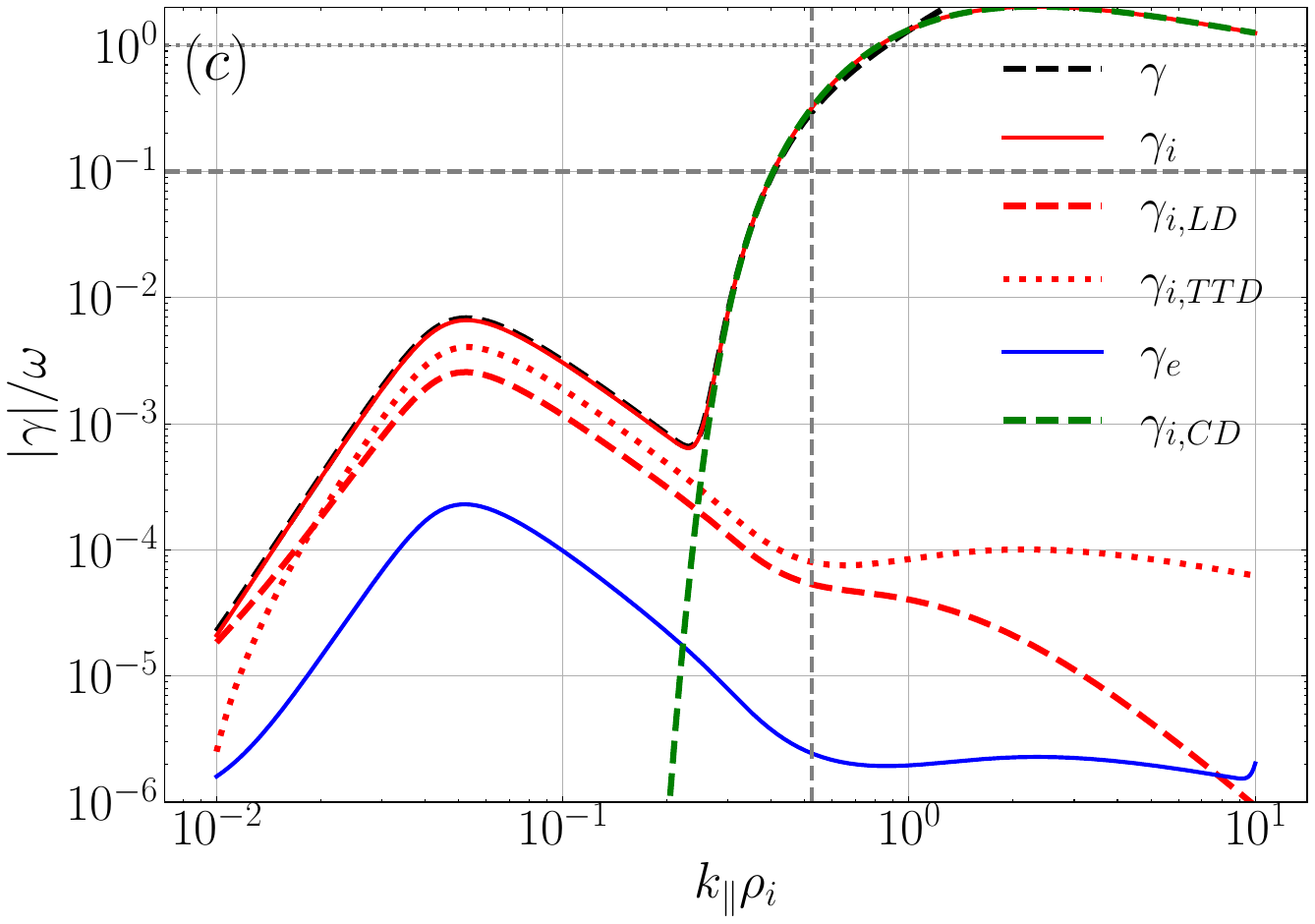}
    \caption{Linear dispersion relation solution for an ICW from \texttt{PLUME} for plasma parameters $\beta_i = 1$, $T_i/T_e = 1$, $v_{ti}/c = 10^{-4}$, and $m_i / m_e = 1836$. We set the perpendicular component of the wavenumber to $k_\perp \rho_i = 0.01$ and present solutions for (a) normalized wave frequency $\omega/\Omega_i$ (black), (b) electric field polarization $\mathcal{P}_E$, and (c) normalized damping rates $|\gamma|/\omega$ as a function of $k_\parallel \rho_i$.  The contributions of ion Landau damping (iLD), ion transit-time damping (iTTD), and ion cyclotron damping (iCD) are separately plotted in (c), showing the rapid and strong onset of ion cyclotron damping as $k_\parallel \rho_i \rightarrow 1$. An empirical analytical expression for the cold plasma ICW wave frequency $\omega/\Omega_i$ (red dashed), given by Eq.~(\ref{eq:cp_icw}), is plotted in (a) for comparison.}
    \label{fig:icw_mode}
\end{figure}

For the parameter set $\beta_i = 1$, $T_i/T_e = 1$, $v_{ti}/c = 10^{-4}$, $m_i / m_e = 1836$, and $k_\perp \rho_i = 0.01$, \texttt{PLUME} scans over the range $10^{-2} \le k_\parallel \rho_i \le 10^1$ to obtain the linear dispersion relation shown in \figref{fig:icw_mode}. We present (a) the wave frequency normalized by the ion cyclotron frequency $\omega/\Omega_i$, (b) the electric field polarization $\mathcal{P}_E$, and (c) the normalized damping rates $|\gamma|/\omega$ as functions of $k_\parallel \rho_i$.
In \figref{fig:icw_mode}(a), we compare the numerical results for the normalized wave frequency with an analytical approximation (red dashed) based on the cold plasma ICW dispersion relation \cite{squire2022high}, here expressed as a function of $k_\parallel \rho_i$ and $\beta_i$
\begin{equation}
    \frac{\omega}{\Omega_i} = \frac{k_\parallel \rho_i}{2 \sqrt{\beta_i}} \left[\sqrt{\left(\frac{k_\parallel \rho_i}{\sqrt{\beta_i}}\right)^2 + 4} - \frac{k_\parallel \rho_i}{\sqrt{\beta_i}} \right]\label{eq:cp_icw}
\end{equation}
The numerical results show good agreement with this approximation for  $k_\parallel \rho_i \ll 1$, where damping remains weak ($-\gamma/\omega < 0.1$).
The electric field polarization plot in \figref{fig:icw_mode}(b) indicates left-hand circular polarization $\mathcal{P}_E \simeq -1$ for $k_\parallel \rho_i > 0.2$, confirming that this is the ICW solution.
The damping rates in \figref{fig:icw_mode}(c) reveal that ion damping dominates over electron damping over the entire range, with ion cyclotron damping ($\gamma_{i,CD}$, green dashed) dominating for $k_\parallel \rho_i \geq 0.2$; we observe that significant ion cyclotron damping ($-\gamma/\omega \geq 0.1$) is expected for $k_\parallel \rho_i \geq 0.4$.

To investigate ion cyclotron damping, we select the ICW mode with $k_\parallel \rho_i = 0.525$ and $k_\perp \rho_i = 0.01$ to construct the electromagntic field eigenfunctions for a single, forward-propagating ICW using Eqs.~(\ref{eq:constructed_em_fields}) and~(\ref{eq:em_ift}). The $k_\parallel \rho_i = 0.525$ mode is marked by the vertical dashed line in \figref{fig:icw_mode}, which exhibits left-handed polarization and a total damping rate of $-\gamma/\omega = 0.3$, indicating that it is a strongly damped ICW. The corresponding wave frequency is $\omega / \Omega_i = 0.2665$, yielding a normalized wave period of $T \Omega_i = 23.58$.
We assign a random phase $\alpha = 0$ to this mode, and set the amplitude scaling factor to $\epsilon = 0.02$.
The Poynting flux for this ICW is primarily along the mean magnetic field in the $+\hat{z}$ direction,
with only 1\% of the energy flux perpendicular to the mean field. 

The electromagnetic eigenfunction for a single, backward-propagating ICW can be generated by choosing all of same parameters except changing the sign of of $k_\parallel \rho_i = -0.525$, yielding a wave Poynting flux primarily in the $-\hat{z}$ direction.  The dispersion characteristics in \figref{fig:icw_mode} remain unchanged under this sign change, and thus are not shown again. 

Finally, we generate a standing ICW wave pattern by the linear superposition of these two counterpropagating ICWs with the same amplitude and the same phase $\alpha = 0$ for both modes, denoted the counterpropagating ICW case.

\subsection{\label{sec:cd_1w_2w} Velocity-Space Signatures of the Damping of Single and Counterpropagating Ion Cyclotron Waves}
Here we apply the Liouville mapping process to generate the velocity-space signatures of ion cyclotron damping in the forward-propagating, backward-propagating, and counterpropagating ICW cases, as defined in Section~\ref{sec:icw_properties}.
For all cases, we set the physical initial time to $t_\text{init} = -4 T$. The FPC is computed at the spatial point $\mathbf{r}_0/\rho_i = (0.1, 0.1, 0.1) $ and the physical final time $t_0 = T$ over the time interval from $0$ to $2T$ with a time step of $\Delta t = 0.025T$. Again, although $\mathbf{r}_0$ is explicitly specified, the system is spatially homogeneous, so the spatial location does not affect the features of the FPC signatures for the propagating wave cases.
Velocity-space grids are set in Cartesian coordinates $(v_x, v_y, v_z)$, with each dimension spanning from $-4 v_{ti}$ to $4 v_{ti}$ using 40 sampled points.

\begin{figure*}
 \begin{center}
\includegraphics[width=1.0\textwidth]{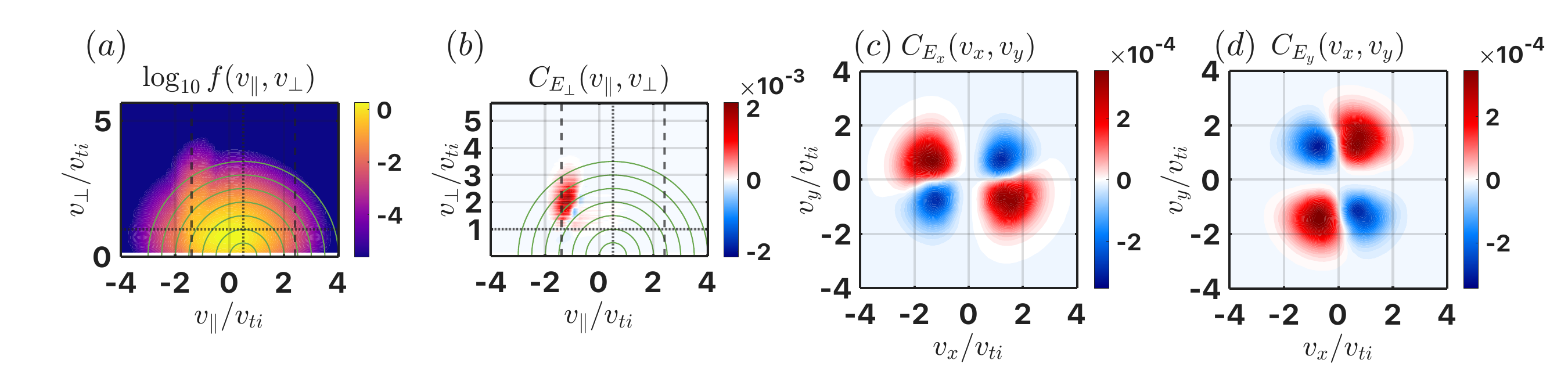}
\includegraphics[width=1.0\textwidth]{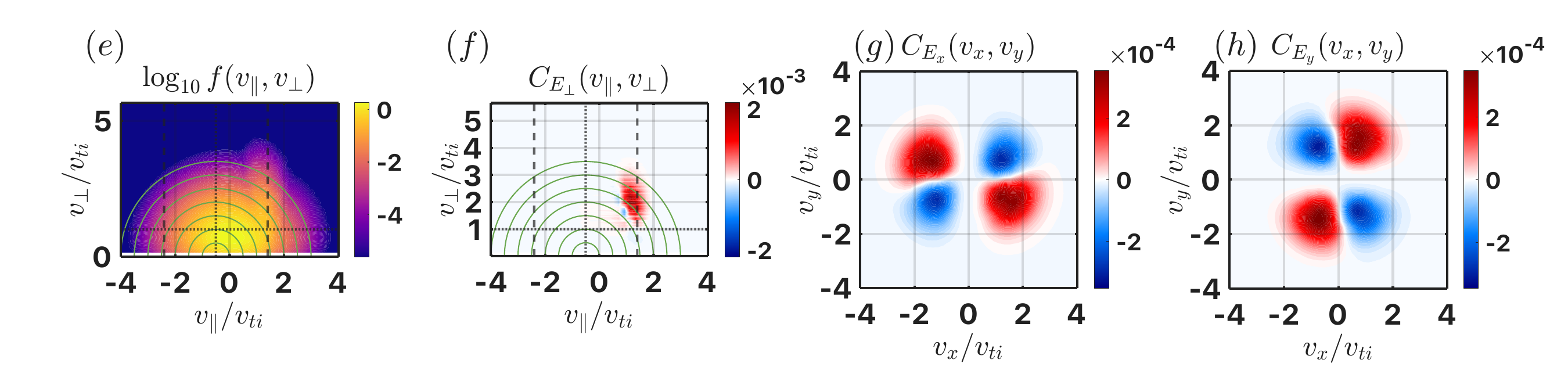}
\includegraphics[width=1.0\textwidth]{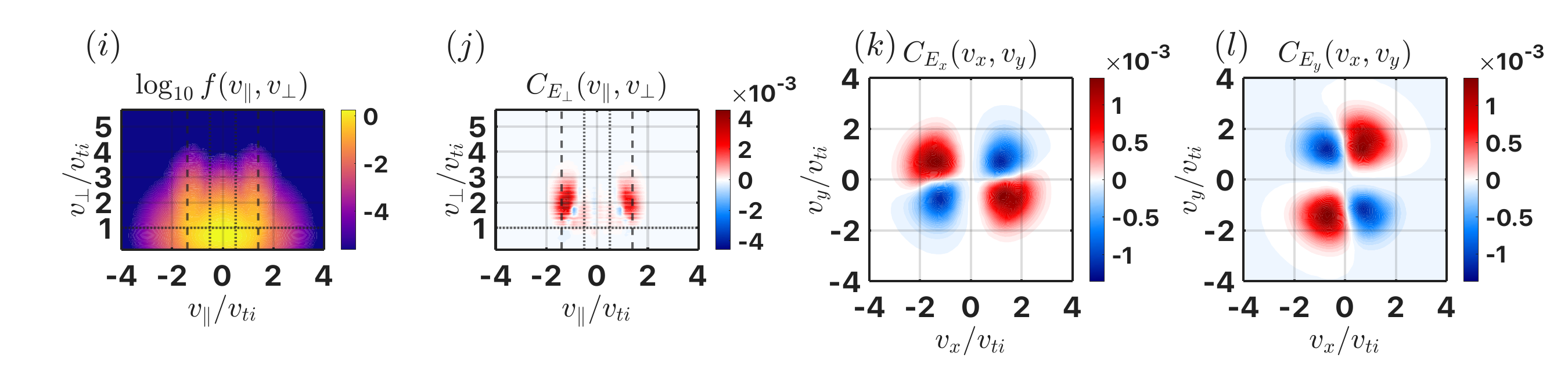}
   \end{center}
\caption{\label{fig:cd_1icw_2icw} From our Liouville mapping technique, plots of the (first column) ion velocity distribution function $f(v_\parallel, v_\perp)$ at the final time slice, (second column) gyrotropic perpendicular FPC $C_{E_\perp}(v_\parallel, v_\perp)$, (third and fourth columns) FPCs reduced to the perpendicular plane $C_{E_x}(v_x, v_y)$ and $C_{E_y}(v_x, v_y)$ for the (first row, a--d) forward-propagating ICW with $k_\parallel \rho_i = 0.525$, (second row, e--h) backward-propagating ICW with $k_\parallel \rho_i = -0.525$, and (third row, i--l) counterpropagating ICWs with $k_\parallel \rho_i = \pm 0.525$.}
\end{figure*}

In \figref{fig:cd_1icw_2icw}, we present the results of the Liouville mapping for the (a--d) forward-propagating ICW, (e--h) backward-propagating ICW, and (i--l) counterpropagating ICW cases, one in each row.  Each column presents (a,e,i) the ion velocity distribution function at the final time slice, (b,f,j) the gyrotropic perpendicular FPC $C_{E_\perp}(v_\parallel, v_\perp)$, and the correlations reduced to the perpendicular plane of velocity space (c,g,k)  $C_{E_x}(v_x, v_y)$ and (d,h,l) $C_{E_y}(v_x, v_y)$.

In all cases, FPCs on the perpendicular plane $C_{E_x}(v_x, v_y)$ and $C_{E_y}(v_x, v_y)$ exhibit nearly identical quadrupolar signatures, consistent with those first discovered through the analysis of \emph{MMS} observations of ion cyclotron damping in Earth's turbulent magnetosheath \cite{afshari2024direct}.  This pattern arises due to the phase difference between the perpendicular electric field components and the ion bulk velocity in the ICW, as discussed in detail in Section~\ref{sec:cd_cexcey_interpret}.

The plots of the distribution function $f(v_\parallel,v_\perp)$ and the perpendicular FPC  $C_{E_\perp} (v_\parallel, v_\perp)$ on the gyrotropic plane, however, are different for these three cases. The normalized parallel wave phase velocity $\omega / (k_\parallel v_{ti})$ of the ICW is labeled with the vertical black dotted line in the gyrotropic plots in the first two columns.  The two resonant parallel velocities, solutions of Eq.~(\ref{eq:res_condition}) for the $n = \pm 1$ resonant modes, $v_{\text{res}, n = \pm 1} / v_{ti}$ shown in the gyrotropic plots for the single wave cases, are marked by the two vertical black dashed lines.  For the gyrotropic plots of the counterpropagating ICW cases in panels (i) and (j), we plot with vertical black dashed lines only the $n=1$ resonant velocities for each of the two modes.

For the forward-propagating ICW case, the wave parallel phase velocity is $\omega / (k_\parallel v_{ti}) = 0.508$, with resonant parallel velocities $v_{\text{res}, n = -1} / v_{ti} = 2.413$ and $v_{\text{res}, n = 1} / v_{ti} = -1.400$.
At the final time plotted in \figref{fig:cd_1icw_2icw}(a), the distribution function shows a shoulder-like extension near the $n = 1$ resonant velocity, as expected theoretically. In contrast, no significant feature appears at the $n = -1$ resonance. This is because cyclotron resonance occurs only when the polarization of the particle's gyromotion matches that of the wave's electric field. In this case, the particles are ions, whose cyclotron motion is left-handed, and the wave is an ICW, which is also left-hand polarized. 
In the gyrotropic plot of the perpendicular FPC $C_{E_\perp} (v_\parallel, v_\perp)$ in \figref{fig:cd_1icw_2icw}(b), we find a net change of the phase-space energy density associated with the $n=1$ resonance at $v_{\text{res}, n = 1} / v_{ti} = -1.400$, showing a region with a gain of phase-space energy density (red) at $v_\perp > v_{ti}$ above a region of weaker loss of phase-space energy density (blue) at $v_\perp < v_{ti}$. 
These features represent the time-averaged acceleration of ions in the perpendicular direction due to the perpendicular electric field, as explored in more detail in Section~\ref{sec:cd_ceperp_interpret}.

The backward-propagating ICW case yields a velocity distribution function $f(v_\parallel,v_\perp)$ and perpendicular FPC  $C_{E_\perp} (v_\parallel, v_\perp)$ on the gyrotropic plane that looks identical to the forward-propagating ICW case but with the sign of $v_\parallel$ reversed, as would be expected theoretically. Here the parallel wave phase velocity is $\omega / (k_\parallel v_{ti}) = -0.508$, and the resonant velocity for the $n = 1$ mode shifts to $v_{\text{res}, n = 1} = 1.400$. For the counterpropagating ICW case, the  features of the FPCs $C_{E_x} (v_x, v_y)$ and $C_{E_y} (v_x, v_y)$ on the perpendicular plane are qualitatively similar to the single ICW cases, but with double the amplitude since the two waves are effectively linearly superposed. The gyrotropic perpendicular FPC $C_{E_\perp}(v_\parallel, v_\perp)$ in \figref{fig:cd_1icw_2icw}(j) displays patterns at both $v_z/v_{ti} = -1.400$ and at $v_z/v_{ti} = 1.400$, corresponding to the $n = 1$ resonant velocities of the forward-propagating and backward-propagating ICWs that both contribute to the energization.  Note that this velocity-space signature of ion energization is dependent on spatial position for the standing wave pattern, and here we show the result at the position of the anti-node in that pattern. For a detailed discussion of the spatial dependence, see Appendix~\ref{sec:icd_asymmetry}.

\subsection{\label{sec:cd_cexcey_interpret} The Quadrupolar Structures of $C_{E_x}(v_x, v_y)$ and $C_{E_y}(v_x, v_y)$}

\begin{figure}
 \begin{center}
 \includegraphics[width=.5\textwidth]{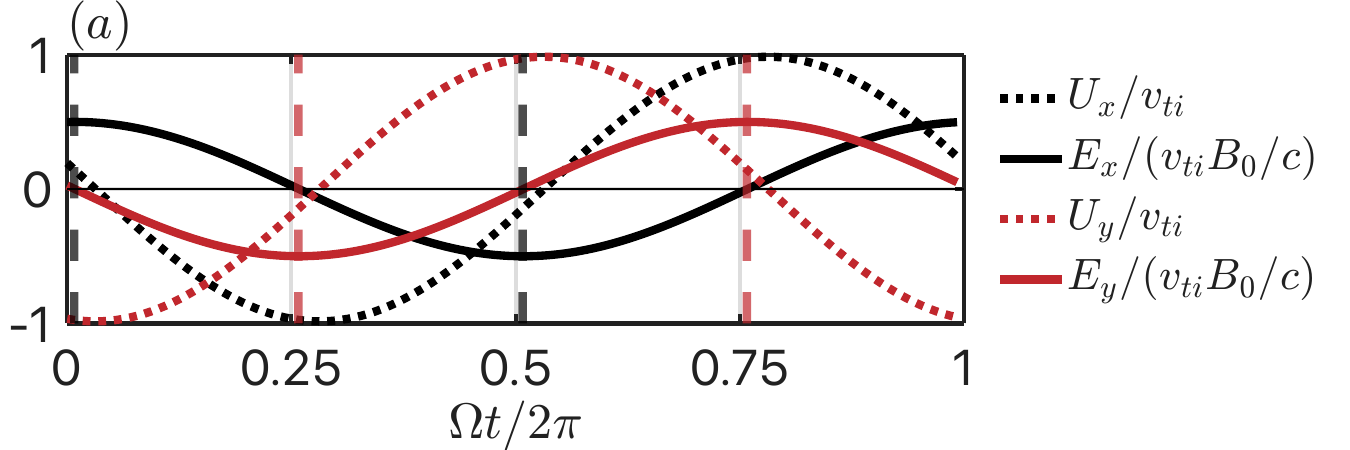}
  \includegraphics[width=.5\textwidth]{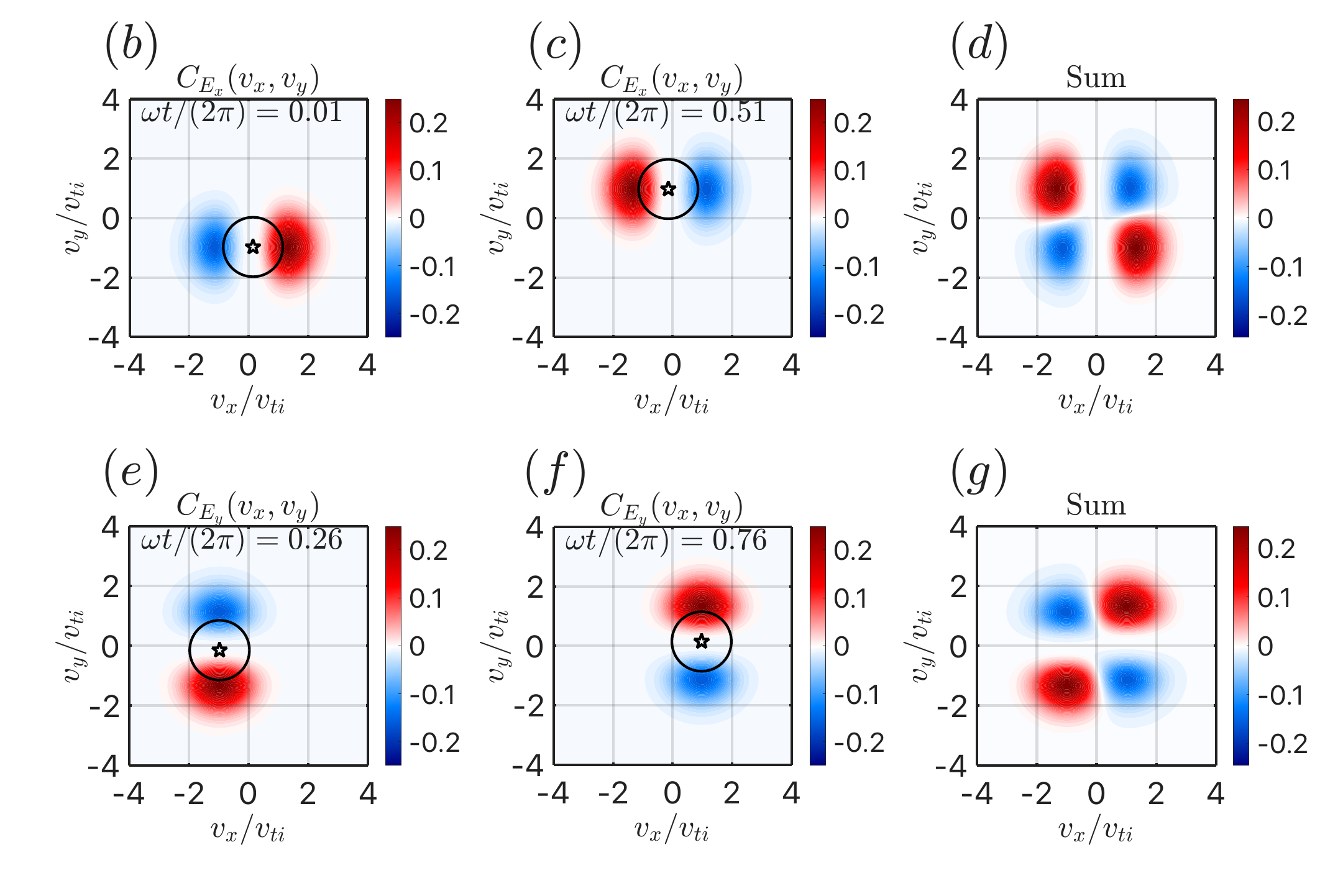}
   \end{center}
\caption{\label{fig:interpret_ceperpVxVy_cd} From \texttt{PLUME} solutions of the ICW eigenfunction, we plot (a) the perpendicular bulk flow velocity components $U_x/v_{ti}$ (black dotted) and $U_y/v_{ti}$ (red dotted) and the perpendicular electric field components $E_x/(v_{ti} B_0/c)$ (black solid) and 
$E_y/(v_{ti} B_0/c)$ (red solid).  The dominant timeslices of $C_{E_x}$ at (b)  $\omega t / (2 \pi) = 0.01$  and (c) $0.51$ combine to produce (d) the quadrupolar velocity-space signature of $C_{E_x}$, and the  timeslices of $C_{E_y}$ at (e)  $\omega t / (2 \pi) = 0.26$  and (f) $0.76$ combine to produce (g) the quadrupolar velocity-space signature of $C_{E_y}$.}
\end{figure}

The quadrupolar structures of $C_{E_x}(v_x, v_y)$ and $C_{E_y}(v_x, v_y)$ can be understood using a simplified model based on the linear eigenfunctions computed by \texttt{PLUME}. Focusing on the example of the forward-propagating single-ICW field with $k_\parallel \rho_i = 0.525$, we approximate the ion distribution function as a Maxwellian centered at the bulk flow velocity $(U_x, U_y)$, obtained from the inverse Fourier transform of the \texttt{PLUME} outputs. This model neglects $v_z$ dependence, effectively focusing on the particles with parallel velocities that are resonant with the wave. 
The model captures the essential dynamics in the perpendicular velocity plane, where $(U_x, U_y)$ follows a clockwise circular trajectory around the origin, corresponding to the left-hand circular polarization of the ICW. The evolution of $(U_x, U_y)$  (dotted)  and $(E_x, E_y)$  (solid) over a full wave period from the \texttt{PLUME} solution for the ICW are plotted in \figref{fig:interpret_ceperpVxVy_cd} (a).  For the discussion below, we denote the distinct phase angles as $\phi_{Q}$ for each of these four fields, where $Q = U_x, E_x, U_y, E_y$.

To estimate $C_{E_x}(v_x, v_y)$, consider that its integrand consists of a velocity-space dependent term, $- v_x^2 \partial f / \partial v_x$ and a velocity-space independent term, $E_x$. For a Maxwellian distribution centered at $(U_x, U_y)$, the velocity-dependent term produces a bipolar structure: negative to the left of $U_x$, positive to the right, and zero at the center. When multiplied by $E_x$, this results in a blue-red bipolar signature (from left to right) centered at $U_x$ for $E_x>0$ or a red-blue bipolar signature for $E_x<0$, as shown in \figref{fig:interpret_ceperpVxVy_cd}(b) at $\omega t / (2 \pi) = 0.01$ and (c) at $\omega t / (2 \pi) = 0.51$. Here we have set the amplitude by specifying  $\epsilon = 0.5$ for ease of visualization. The same reasoning applies to $C_{E_y}$ with respect to $v_y$ and $E_y$, shown in \figref{fig:interpret_ceperpVxVy_cd}(e) and (f).

The quadrupolar patterns in $C_{E_x}$ and $C_{E_y}$, shown in \figref{fig:cd_1icw_2icw} (c) and (d), arise from the phase relationships between the ion bulk flow and the electric field components, $\phi_{U_x}$, $\phi_{U_y}$, $\phi_{E_x}$ and $\phi_{E_y}$. To visualize this, we examine $C_{E_x}$ at four evenly spaced time slices, $\omega t / (2 \pi) = 0.01, 0.26, 0.51, 0.76$, during the wave period. Among these, the two slices where $E_x$ is at its peak and trough dominate the time-averaged response, since the contributions from the other two selected time slices are zero because $E_x = 0$ at those times. For $C_{E_x}$, these two dominant snapshots are marked in \figref{fig:interpret_ceperpVxVy_cd}(a) with vertical black dashed lines.
As shown in the second row of \figref{fig:interpret_ceperpVxVy_cd}, at each time slice, the bulk perpendicular flow velocity $(U_x/v_{ti}, U_y/v_{ti})$ is marked with a star, while a surrounding circle with a radius of one thermal velocity represents the ion distribution along its clockwise gyrating motion around the velocity-space origin.
At $\omega t / (2 \pi) = 0.01$, where $E_x > 0$, the resulting blue-red dipole appears in the lower half of the $(v_x, v_y)$ plane with $U_x \gtrsim 0$ and $U_y < 0$.
At $\omega t / (2 \pi) = 0.51$, where $E_x < 0$, the pattern reverses, appearing as a red-blue dipole in the upper half of the plane with $U_x \lesssim 0$ and $U_y > 0$.
Summing these two dominant patterns yields the composite shown in \figref{fig:interpret_ceperpVxVy_cd}(d). This sum effectively represents the time-averaged $C_{E_x}$ over one wave period and reproduces qualitatively the quadrupolar pattern seen in \figref{fig:cd_1icw_2icw}(c).

Analogously for $C_{E_y}$, the two dominant time slices now occur at $\omega t / (2 \pi) = 0.26$ and~$0.76$, as labeled by vertical red dashed lines in \figref{fig:interpret_ceperpVxVy_cd}(a). The red-blue (from bottom to top) bipolar signature of $C_{E_y}$ at $\omega t / (2 \pi) = 0.26$ is shown in 
 \figref{fig:interpret_ceperpVxVy_cd}(e) and the blue-red  signature  at $\omega t / (2 \pi) = 0.76$  in (f), with their summing in (g) clearly qualitatively reproducing the quadrupolar pattern of $C_{E_y}$ seen in \figref{fig:cd_1icw_2icw}(d).
 
In summary, the quadrupolar structures of $C_{E_x}(v_x, v_y)$ and $C_{E_y}(v_x, v_y)$ originate from phase differences between the electric field and the ion bulk flow velocity components, which is evident by approximating the evolving distribution as a Maxwellian centered at $(U_x, U_y)$. Note that this model neglects distortions in $(v_x, v_y)$ plane and any dependence on $v_z$. Next, we explore the variation of the gyrotropic FPC  $C_{E_\perp}(v_\parallel, v_\perp)$ as a function of $v_z$.

\subsection{\label{sec:cd_ceperp_interpret} The Ion Energization in the Gyrotropic Velocity-Space Signature $C_{E_x}(v_\parallel, v_\perp)$}
\begin{figure}
 \begin{center}
 \includegraphics[width=.5\textwidth]{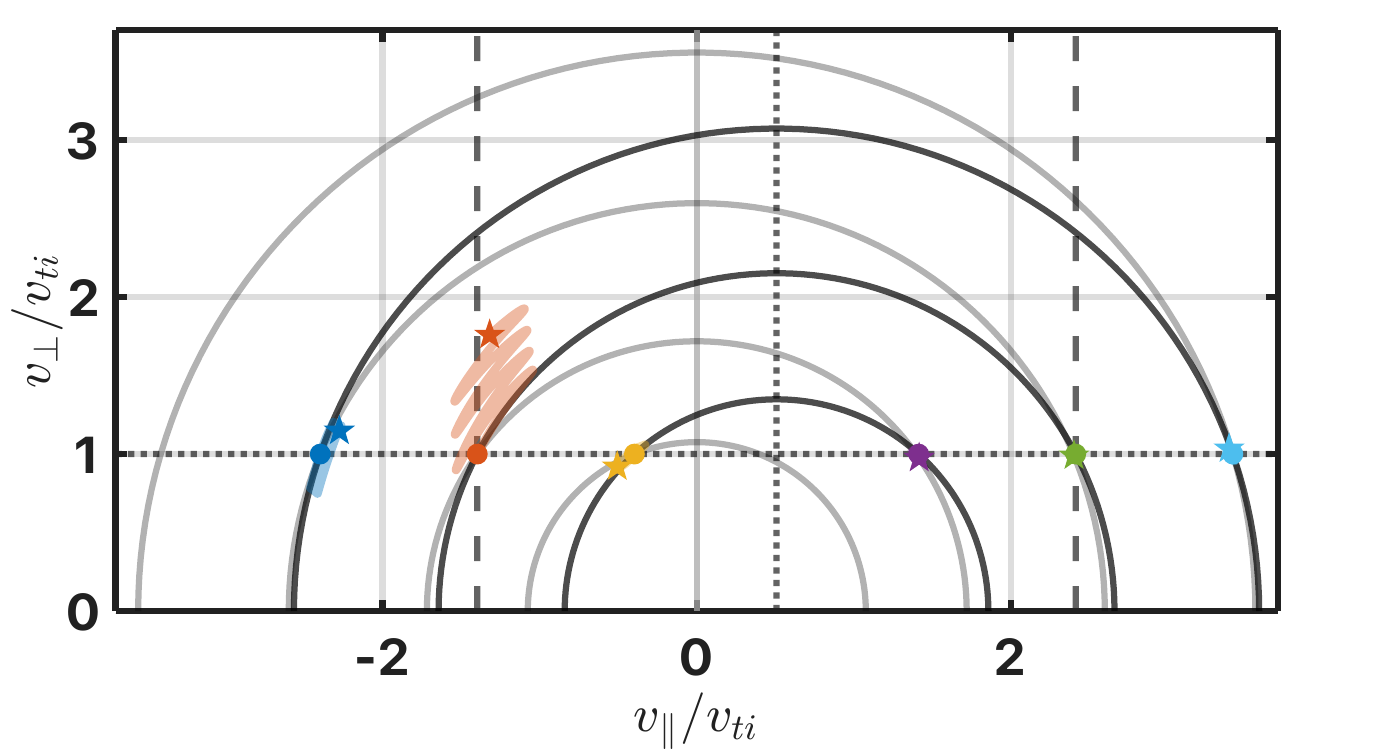}
   \end{center}
\caption{\label{fig:interpret_ceperpVperpVz_cd} Selected single ion trajectories in the forward-propagating ICW fields with $k_\parallel \rho_i = 0.525$ on the gyrotropic  $(v_\parallel, v_\perp)$ plane. The vertical black dotted line labels the wave parallel phase velocity $\omega/(k_\parallel v_{ti}) = 0.508$, and the two vertical black dashed lines mark the two resonant velocity $v_{\text{res}, n = -1} / v_{ti} = 2.413$ and $v_{\text{res}, n = 1} / v_{ti} = -1.400$.  Initial velocites (dots) and final velocities (stars) are indicated. }
\end{figure}

To understand the features of the gyrotropic FPC $C_{E_\perp}(v_\parallel, v_\perp)$, we examine single particle motion trajectories in the gyrotropic plane  for the forward-propagating ICW. In \figref{fig:interpret_ceperpVperpVz_cd}, we  show six such ion trajectories, each evolving over six wave periods. All ions start from the same spatial position $\mathbf{r} = (0.1, 0.1, 0.1) \rho_i$, with identical initial perpendicular velocity $v_{x0}/v_{ti} = 1$ and $v_{y0}/v_{ti} = 0$. Their initial parallel velocities are: $v_{z0}/v_{ti} = -2.400$ (blue), $-1.400$ (red), $-0.400$ (yellow), $ 1.413$ (purple), $ 2.413$ (green), and $ 3.413$ (cyan). The initial parallel velocities of the red and green trajectories correspond to the $n = 1$ and the $n = -1$ resonant velocities (vertical dashed lines), respectively. Dots and stars mark their initial and final velocity-space positions on the gyrotropic plane, respectively.

To visualize how the energy of the ions evolves in the ICW  fields, we overlay two sets of semicircles in \figref{fig:interpret_ceperpVperpVz_cd}: gray contours represent constant energy in the lab frame, while black contours represent constant energy in the frame moving at the parallel phase velocity of the ICW (hereafter referred as the ICW frame). Comparing initial and final velocity-space positions reveals that only the red trajectory, starting at the $n = 1$ resonant velocity ($v_{z0}/v_{ti} = v_{\text{res}, n=1}/v_{ti} = -1.400$), experiences significant energy gain. The other five trajectories remain close to their respective initial energy contours.

All ions undergo left-hand polarized cyclotron motion around the background magnetic field $\V{B}_0= B_0 \hat z$. Although this lowest-order motion is not visible in the gyrotropic plane as the azimuthal angle is averaged out, it does not contribute to energy gain and therefore does not affect our analysis of energization. The higher-order dynamics, which are responsible for net energy gain, are captured in the red trajectory. 

The first of these is pitch angle scattering, where the particle moves back and forth along the black semicircle. In the ICW frame, this motion redistributes kinetic energy between parallel and perpendicular components without changing the total energy. In the lab frame, however, the total energy increases as the pitch angle in the ICW frame decreases (moving the particle outward relative to the gray, constant energy contours in the lab frame), and vice versa. 
Given a Maxwellian velocity distribution centered at the origin in the lab frame, more particles tend to occupy large pitch angles in the ICW frame than small ones. As a result, more particles gain energy than lose it, leading to a net transfer of energy from the electromagnetic fields to the particles. 

The red trajectory also shows a slow outward drift perpendicular to the energy contours in the ICW frame. Unlike pitch-angle scattering, which does not result in energy gain in the ICW frame and only leads to energization in the lab frame by relying on the fact that the number of particles gaining energy exceeds those losing energy, this outward drift represents an energy gain in both the ICW and lab frames for each ion individually.

Together, pitch-angle scattering and outward drift to higher energy result in a net energy gain along the positive $v_\perp$ direction. This is the key dynamics underlying the upward red lobe in $C_{E_\perp}(v_\parallel, v_\perp)$, as captured by FPC  shown in \figref{fig:cd_1icw_2icw}(b). This indicates an increasing  probability of finding particles after the system has evolved for $5T$ with $v_\parallel \sim v_{res, n=1}$ and $v_\perp > v_{ti}$, and the consequent decreasing probability for particles with $v_\parallel \sim v_{res, n=1}$ and $v_\perp < v_{ti}$ (required for conservation of particle number).

One may notice that the $C_{E_\perp}(v_\parallel, v_\perp)$ patterns reported here, \figref{fig:cd_1icw_2icw}(b, f), differ from those shown in Fig. 4(a) of the previously work Klein \emph{et al.} (2020) \cite{klein2020diagnosing}. This difference arises because Klein \emph{et al.} (2020) \cite{klein2020diagnosing} analyzed broadband turbulence containing many wave modes, each associated with its own resonant velocity. Around each resonance, one expects localized structures, i.e. a red lobe above $v_\perp/v_{ti}\sim 1$ and a faint blue lobe below, as seen in our \figref{fig:cd_1icw_2icw} (b, f). When many such modes are present, their individual signatures overlap, producing the broader, blended pattern observed in Klein \emph{et al.} (2020) \cite{klein2020diagnosing}.

In summary, the velocity-space signature of ion cyclotron damping is characterized by the perpendicular gyrotropic velocity-space signature $C_{E_\perp} (v_\parallel, v_\perp)$ in \figref{fig:cd_1icw_2icw}(b) along with the two perpendicular  plane signatures  $C_{E_x}(v_x, v_y)$ and  $C_{E_y}(v_x, v_y)$ in \figref{fig:cd_1icw_2icw}(c) and~(d). These signatures are consistent with the previous observational identification using \emph{MMS} observations in Earth's turbulent magnetosheath plasma \citep{afshari2024direct}.

\subsection{\label{sec:cd_beta} Variation of the Velocity-Space Signature of Ion Cyclotron Damping with $\beta_i$}
\begin{figure*}
 \begin{center}
\includegraphics[width=.48\textwidth]{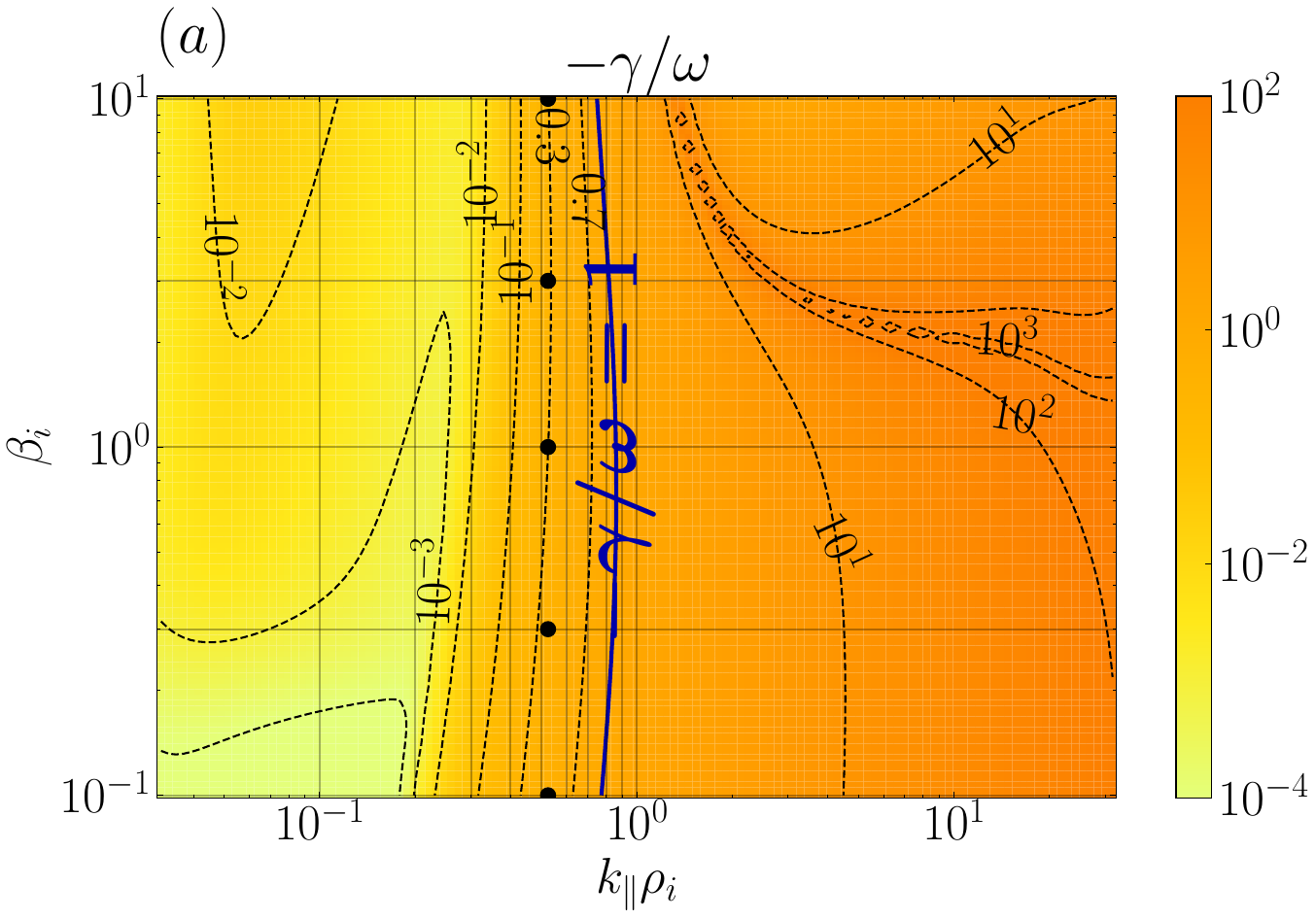}
\includegraphics[width=.48\textwidth]{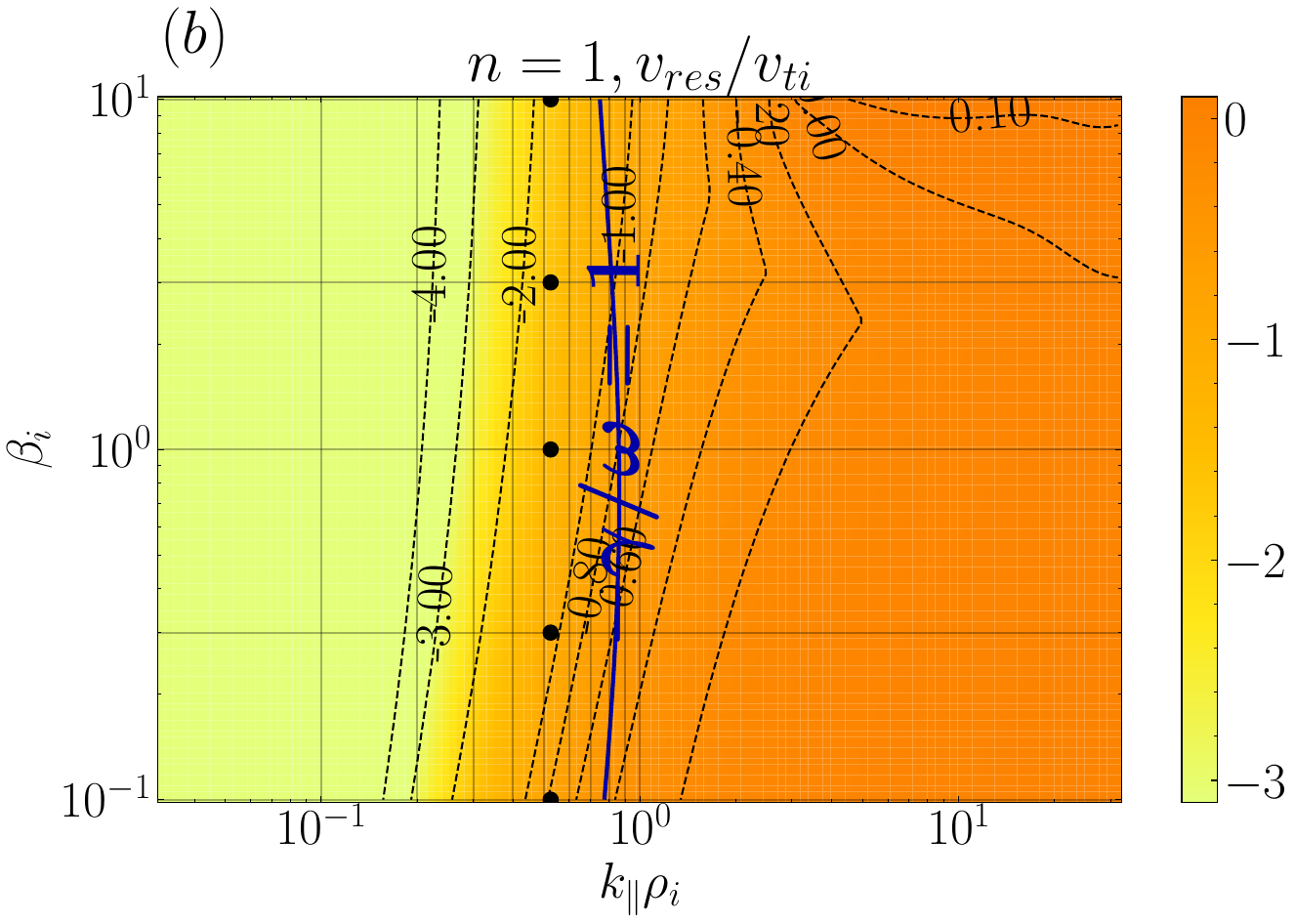}
\includegraphics[width=.48\textwidth]{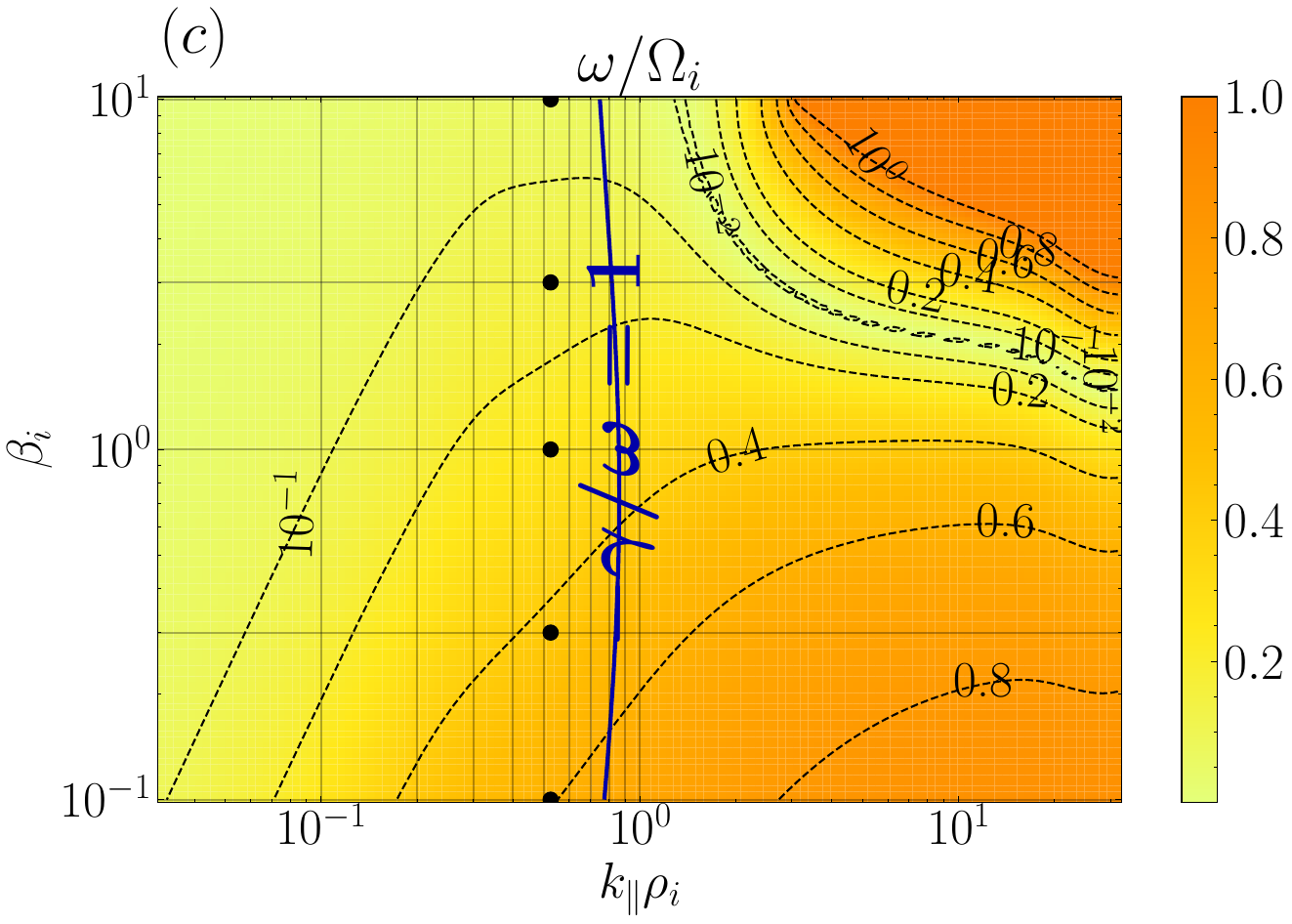}
\includegraphics[width=.48\textwidth]{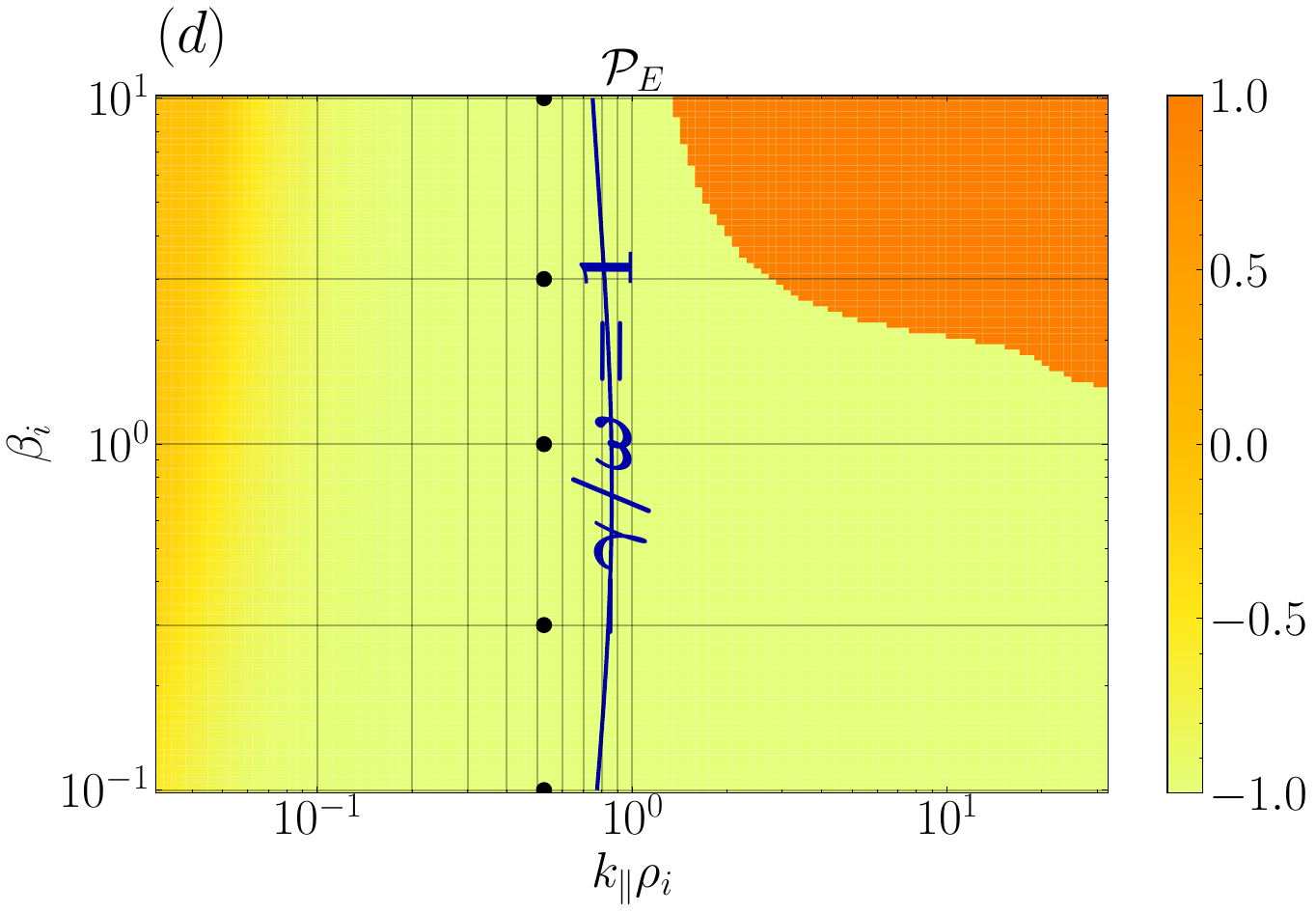}
   \end{center}
 \caption{\label{fig:cd_chosen_modes_beta} For plasma parameters  $T_i/T_e = 1$, $v_{ti}/c = 10^{-4}$ and $m_i / m_e = 1836$ and perpendicular wavenumber  $k_\perp \rho_i=0.01$,   \texttt{PLUME} solutions on the  $(k_\parallel \rho_i,\beta_i)$ parameter space for the (a) normalized damping rate $-\gamma / \omega$, (b) $n = 1$ mode resonant velocity $v_{\text{res}, n = 1}/v_{ti}$, (c)  normalized wave frequency $\omega/\Omega_i$, and (d)  electric field polarization $\mathcal{P}_E$. The chosen ICW modes at $\beta_i = 0.1, 0.3, 1, 3, 10$ are labeled as five black dots on each panel.}
\end{figure*}

Here we use Liouville mapping to investigate how the velocity-space signature of ion cyclotron damping varies with  ion plasma beta $\beta_i$. To have an overall sense of the properties of the ICW as $\beta_i$ varies, we use \texttt{PLUME} to solve for the linear dispersion relation of ICWs over the 2D parameter space $(k_\parallel \rho_i,\beta_i)$, covering the ranges $0.1 \sqrt{0.1} \le k_\parallel \rho_i \le 10\sqrt{10}$  and $0.1 \le \beta_i \le 10$. We hold constant the parameters $k_\perp \rho_i = 0.01$, $T_i/T_e = 1$, $v_{ti}/c = 10^{-4}$ and $m_i / m_e = 1836$. 
In \figref{fig:cd_chosen_modes_beta}(a), we plot a colormap of the normalized total damping rate, $-\gamma/\omega$, on a logarithmic scale, with black dashed contours denoting $-\gamma/\omega = 10^{-4}, 10^{-3},\ldots, 10^3$. The $-\gamma / \omega = 1$ contour is highlighted in solid blue, which appears nearly vertical and closely follows the $k_\parallel \rho_i = 0.8$ grid line. This contour effectively divides the parameter space: on the left are more weakly damped ICWs with  $-\gamma / \omega < 1$, and on the right ICWs are extremely strongly damped, decaying within a  fraction of a wave period when $-\gamma / \omega > 1$. Since our Liouville mapping approach assumes constant amplitude (rather than decaying) wave fields, we restrict our application of the technique to parameter choices  within the $-\gamma / \omega < 1$ region.

As discussed in Section~\ref{sec:cd_results}, a key feature of the gyrotropic FPC $C_{E_\perp}(v_\parallel, v_\perp)$ is the localization of the net ion energization near the $n = 1$ resonant velocity. To examine how this resonant velocity varies with $\beta_i$, we plot $v_{\text{res}, n = 1}/v_{ti}$ over the $(k_\parallel \rho_i, \beta_i)$ plane in \figref{fig:cd_chosen_modes_beta}(b). We set the lower limit of the color bar to $-3$, as values with  $v_{\text{res}, n = 1}/v_{ti}<-3$ yield extremely weak ion cyclotron damping rates. This is because the Maxwellian velocity distribution contains very few particles in velocity-space regions far from the origin.

Notably, within the $-\gamma / \omega < 1$ region, the contours of $v_{\text{res}, n = 1}/v_{ti}$ in \figref{fig:cd_chosen_modes_beta}(b) remain nearly vertical, indicating a weak dependence of the resonant velocity on $\beta_i$. To understand this behavior, we rewrite the resonance condition from Eq.~(\ref{eq:res_condition}) to express
$v_{\text{res}, n}/v_{ti}$ in terms of $\beta_i$ and $k_\parallel
\rho_i$,
\begin{equation}
    \frac{v_{\text{res}, n}}{v_{ti}} = \frac{1}{k_\parallel \rho_i} \left( \frac{\omega}{\Omega_i} - n\right)
    \label{eq:res_condition2}
\end{equation}
At first glance, there appears to be no $\beta_i$ dependence in this expression. However, the normalized wave frequency $\omega/\Omega_i$ does depend on $\beta_i$ in a nontrivial way, as shown in \figref{fig:cd_chosen_modes_beta}(c).  Since we are analyzing the $n=1$ resonance condition,  this plot essentially represents the ratio of the first term  ($\omega/\Omega_i$) to the second term ($n = 1$) inside the parentheses in Eq.~(\ref{eq:res_condition2}).
At $\beta_i \gtrsim 1$, the resonant velocity is dominated by second term ($n = 1$), while at $\beta_i <1$, a weak dependence on $\beta_i$ emerges. However, even for the $\beta_i=0.1$ case where $\omega/\Omega_i \simeq 0.65$ crossing the contour of $-\gamma / \omega = 1$ (solid blue line), the second term still dominates (nearly twice of the first), so the effect of the $\beta_i$ dependence remains  weak.

Finally, in \figref{fig:cd_chosen_modes_beta}(d) we plot the electric field polarization for the ICW mode over the  $(k_\parallel \rho_i,\beta_i)$ parameter space, showing a value  $\mathcal{P}_E = -1$ across most of the parameter plane. This plot confirms the left-hand circularly polarized electric field theoretically expected for ICWs.

Given the weak dependence on $\beta_i$ of the ICW properties shown in  \figref{fig:cd_chosen_modes_beta}, we fix $k_\parallel \rho_i = 0.525$ and select five representative $\beta_i$ values,  $\beta_i=0.1, 0.3,1, 3, 10$, marked with black dots in all panels of \figref{fig:cd_chosen_modes_beta}. For these five  parameter choices, 
the corresponding values of normalized damping rate $-\gamma / \omega$, resonant velocity $v_{\text{res}, n = 1}/v_{ti}$, wave frequency $\omega/\Omega_i$, and wave period  $T \Omega_i$  from the \texttt{PLUME} solutions are summarized in Table~\ref{tab:params_beta}. Two important trends emerge from Table~\ref{tab:params_beta}: (i) the normalized ICW mode period, $T \Omega_i$, increases significantly with $\beta_i$; and (ii) the magnitude of $v_{\text{res}, n = 1}/v_{ti}$ only increases modestly, from $-0.777$ at $\beta_i = 0.1$, to $-1.773$ at $\beta_i = 10$.

\begin{figure*}
 \begin{center}
\includegraphics[width=.9\textwidth]{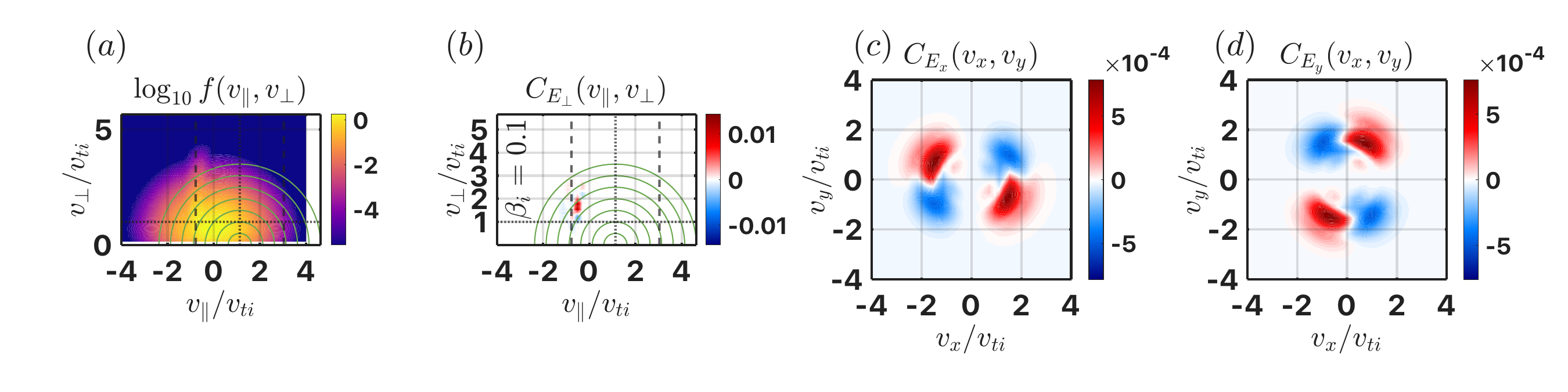}
\includegraphics[width=.9\textwidth]{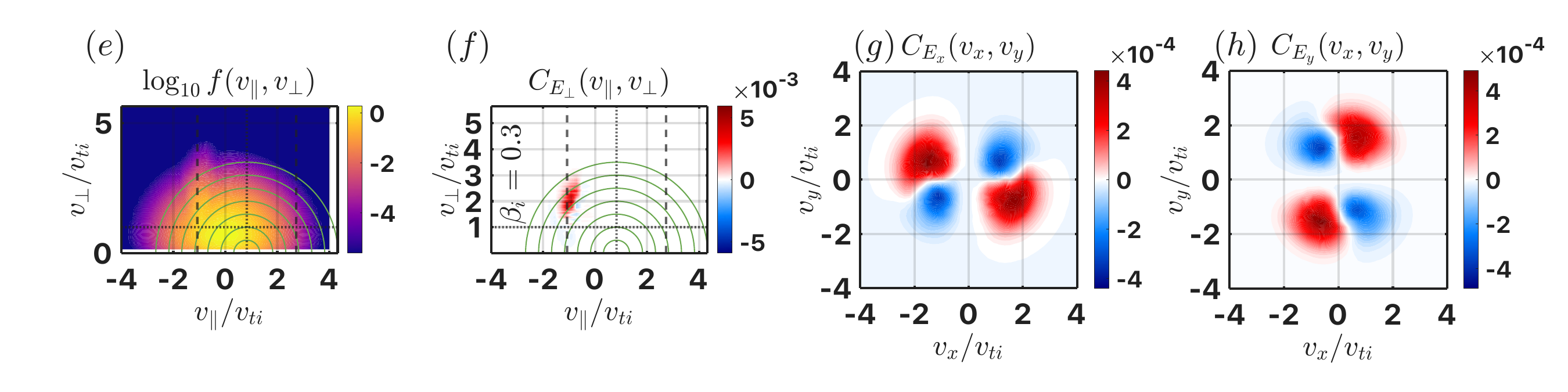}
\includegraphics[width=.9\textwidth]{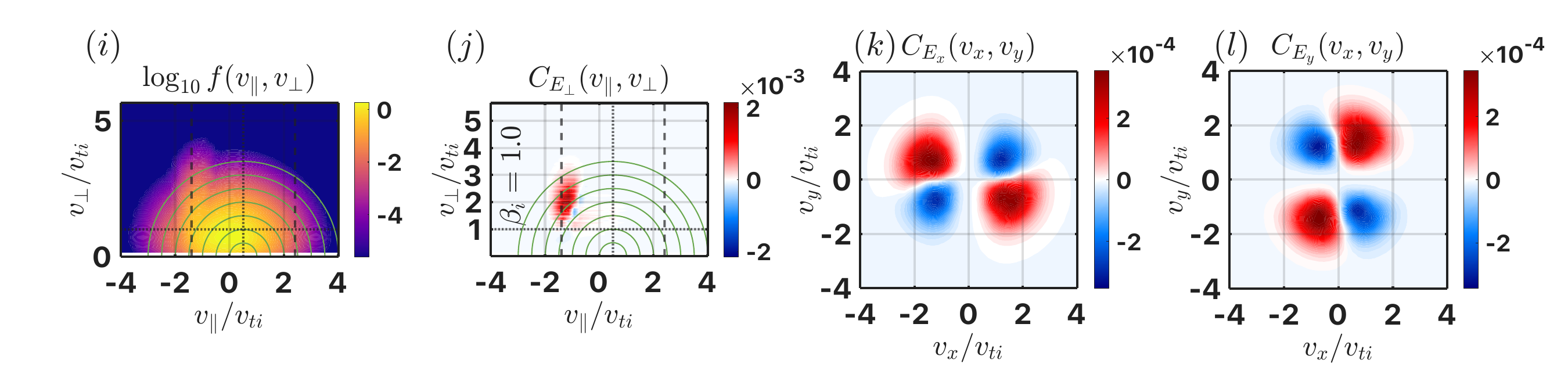}
\includegraphics[width=.9\textwidth]{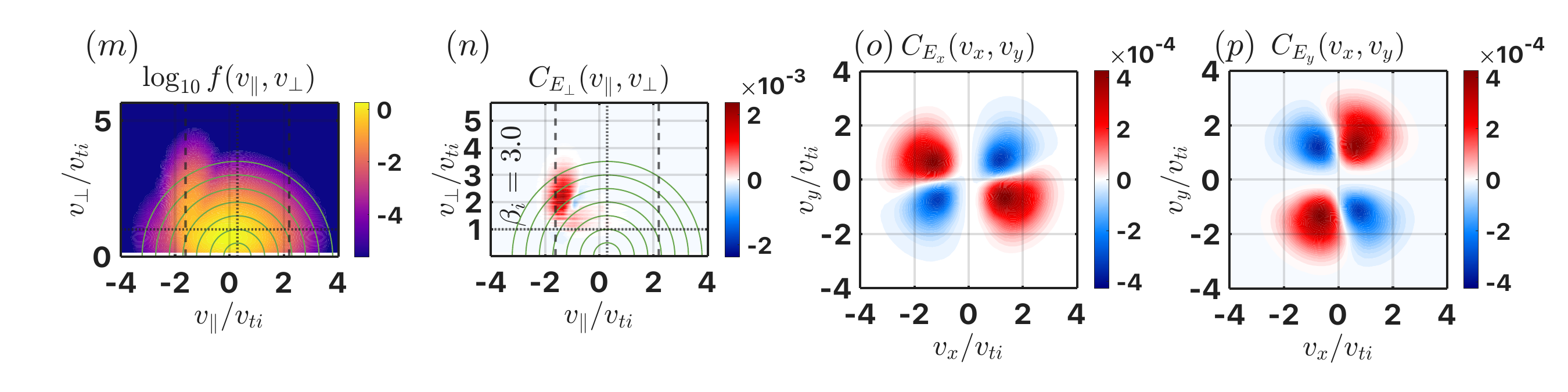}
\includegraphics[width=.9\textwidth]{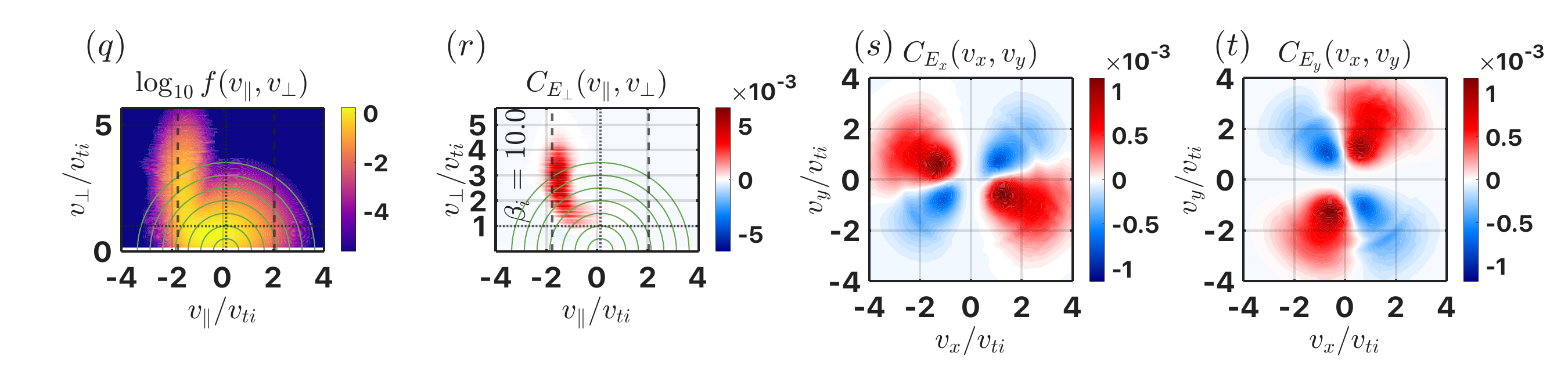}
   \end{center}
 \caption{\label{fig:cd_beta} The  ion velocity distribution function  $f(v_\parallel,v_\perp)$,  gyrotropic perpendicular FPC $C_{E_\perp}(v_\parallel, v_\perp)$, and  FPCs reduced to the perpendicular plane  $C_{E_x}(v_x, v_y)$ and $C_{E_y}(v_x, v_y)$, in the same format as  \figref{fig:cd_1icw_2icw}. From top to bottom, rows present the results for $\beta_i = 0.1, 0.3, 1, 3, 10$. }
\end{figure*}

\begin{table}
\begin{ruledtabular}
\begin{tabular}{ccccc}
 $\beta_i$
& $-\gamma / \omega$ & $v_{\text{res}, n = 1}/v_{ti}$ & $\omega/\Omega_i$ & $T \Omega_i$  \\ \hline
 0.1 & 0.488 & -0.777 & 0.592 & 10.613 \\
 0.3 & 0.374 & -1.081 & 0.433 & 14.526 \\
 1 & 0.298 & -1.400 & 0.267 & 23.575\\
 3 & 0.275 & -1.618 & 0.151 & 41.672\\
 10 & 0.293 & -1.773 & 0.0693 & 90.662\\
\end{tabular}
\end{ruledtabular}
\caption{\label{tab:params_beta} For  $k_\parallel \rho_i=0.525$ and the five $\beta_i$ values, the resulting \texttt{PLUME} solutions for  normalized damping rate $-\gamma / \omega$, resonant velocity $v_{\text{res}, n = 1}/v_{ti}$, wave frequency $\omega/\Omega_i$, and wave period  $T \Omega_i$.}
\end{table}
For the five $\beta_i$ choices in  Table~\ref{tab:params_beta}, we use Liouville mapping to predict the velocity-space signatures of ion cyclotron damping in the same format as \figref{fig:cd_1icw_2icw}.  For all cases, the physical initial time is set to $t_\text{init} = -4T$, and the FPC is computed over a correlation interval $0 < t_f < 2T$.  The resulting velocity-space signatures are presented in \figref{fig:cd_beta}, with the $\beta_i$ in each row increasing from top to bottom.
Although the velocity-space signature reduced to the perpendicular plane  $C_{E_x}(v_x, v_y)$ (third column) and $C_{E_y}(v_x, v_y)$ (fourth column) differ quantitatively (such as amplitude and overall shape of the quadrupolar pattern), the qualitative quadrupolar pattern remains unchanged as  $\beta_i$ varies:  for $C_{E_x}$, red lobes appear in quadrants II and IV; and for $C_{E_y}$, in quadrants I and III. 

For the gyrotropic velocity-space signatures $C_{E_\perp}(v_\parallel, v_\perp)$ (second column), the signature remains well-localized near $v_{\text{res}, n = 1}$ (leftmost vertical dashed line) across all $\beta_i$ cases. In addition, the zero-crossing from loss (faint blue) to gain (red) of phase-space energy density occurs at $v_\perp \sim v_{ti}$, independent of $\beta_i$.


A noticeable trend in  \figref{fig:cd_beta} is that, as $\beta_i$ increases, the distribution function $f(v_\parallel,v_\perp)$ (first column) and gyrotropic velocity-space signature $C_{E_\perp}(v_\parallel, v_\perp)$ (second column) become increasingly extended to higher $v_\perp$ values. This variation is a consequence of our implementation of the Liouville mapping technique and the strong variation of wave period  $T \Omega_i$  as a function of $\beta_i$, as shown in  Table~\ref{tab:params_beta}.
Although the physical initial time is consistently set relative to the wave period to $-4 T$, ensuring each system evolves for four ICW periods before computing the FPC, the absolute value of the normalized ICW mode period $T \Omega_i$ increases significantly with $\beta_i$. At $\beta_i = 0.1$, the system evolves for $4 T \Omega_i = 42.452$ before the computation of FPC on the correlation interval of $\tau \Omega_i = 21.22$; in contrast,  at $\beta_i = 10$, these values increase to $4 T \Omega_i = 362.648$ and $\tau \Omega_i = 181.324$.
We present a direct comparison of cases with matched absolute evolution time and correlation  interval for different $\beta_i$ values in Appendix~\ref{sec:icw_absTime}.

In summary, our general prediction from this study is that the velocity space signatures of ion cyclotron damping do not vary qualitatively with $\beta_i$, except for the relatively weak quantiative change in the $n=1$ resonant velocity, $v_{\text{res}, n = 1}$.

\section{\label{sec:conclu} Conclusion}
Resonant wave-particle interactions are fundamental mechanisms responsible for the collisionless damping of electromagnetic waves in plasmas. The field-particle correlation (FPC) technique has proven to be an effective tool for characterizing these interactions by revealing their distinctive  velocity-space signatures of particle energization. While the $n = 0$ Landau resonance has been extensively studied (including the identification of its velocity-space signatures for both ions \cite{klein2016measuring, howes2017diagnosing, klein2017diagnosing} and electrons \cite{conley2023characterizing}, the physical interpretation of these signatures \cite{klein2016measuring}, their dependence on plasma parameters \cite{conley2023characterizing, huang2024velocity}, and their detection in numerical simulations \cite{Horvath:2020}, spacecraft observations \cite{chen2019evidence, afshari2021importance}, and laboratory experiments \cite{schroeder2021laboratory}), cyclotron damping via the $n = \pm 1$ resonances  remains less throughly explored. This situation is largely due to the challenge of obtaining well-controlled data where cyclotron damping dominates over other energization processes.

This study addresses two key gaps: (1) the lack of a computationally efficient method for generating well-controlled data, and (2) limited understanding of the physics underlying ion cyclotron damping. We fill the first gap using the Liouville mapping method and the second by systematically using Liouville mapping with the FPC technique to characterize the velocity-space signatures of ion cyclotron damping.

The Liouville mapping method is based on Liouville's theorem, which states that the distribution function remains constant along particle trajectories in phase space. In plasma physics, this corresponds to solving the Vlasov equation implicitly by numerically solving the equations of motion for individual particles.

We integrate this approach with the FPC framework using the procedure illustrated in Fig.~\ref{fig:lmfpc_workflow}, with several caveats. The electromagnetic fields are analytically specified from eigenfunctions obtained using solutions to the linear Vlasov-Maxwell dispersion relation. Although we neglect the collisionless damping of the electromagnetic fields of the waves by assuming a fixed peak amplitude in time and ignore the self-consistent distortions to the fields from particle motion, both of which make the system non-self-consistent, we argue in Section~\ref{sec:lm_misc} that this approximation is reasonable for qualitatively predicting the velocity-space signatures of collisionless damping of the waves.

We validate our method by recovering the known velocity-space signature of Landau damping from our Liouville mapping approach with single propagating and standing kinetic \Alfven waves (KAWs), as shown in Fig.~\ref{fig:ld_1kaw_2kaw}. We then apply the technique to ion cycltron waves (ICWs) and investigate the velocity-space signature of ion cyclotron damping in unprecedented detail, including comprehensive interpretations of the quadrupolar features, cross-comparisons with single particle trajectories, and variations with $\beta_i$.

This study yields several key findings about velocity-space signature of ion cyclotron damping. First, ion cyclotron damping produces a quadrupolar pattern in the $(v_x, v_y)$ plane, driven by the phase relations between the ion bulk flow and the perpendicular electric field.  This behavior can be intuitively understood by modeling the motion of the ion distribution as a Maxwellian centered at the bulk flow velocity, as detailed in Fig.~\ref{fig:interpret_ceperpVxVy_cd}. We note that these quadrupolar velocity-space patterns have distinct physical meanings from those reported by \citet{norgren2025phase} and \citet{shuster2021structures}. Second, ion cyclotron damping leads to net ion energization to higher $v_\perp$ near the $n = 1$ resonant parallel velocity in the $(v_\parallel, v_\perp)$ plane, driven by pitch-angle scattering and acceleration to higher energy, both in the frame of the ICW. This process is revealed by examining single-particle trajectories in the $(v_\parallel, v_\perp)$ plane, as visualized in Fig.~\ref{fig:interpret_ceperpVperpVz_cd}. Third, these features remain largely unchanged across a broad range of ion plasma beta values, $\beta_i = 0.1, 0.3, 1, 3, 10$, as shown in Fig.~\ref{fig:cd_beta}. To isolate the influence of $\beta_i$, we fix the parallel wavenumber at $k_\parallel \rho_i = 0.525$. This choice is motivated by the finding that the $n = 1$ resonant velocity depends only weakly on $\beta_i$ and is instead primarily determined by $k_\parallel \rho_i$, as shown in Fig.~\ref{fig:cd_chosen_modes_beta}.

To our knowledge, this work presents the first detailed characterization of how ion cyclotron damping velocity-space signatures vary with $\beta_i$. These results offer valuable benchmarks for identifying ion cyclotron damping in spacecraft observations and provide a methodical foundation for future studies of other wave-particle interactions.


\begin{acknowledgments}
Rui Huang thanks Kristopher G. Klein for invaluable assistance with \texttt{PLUME}, and Collin R. Brown and Daniel J. McGinnis for insightful discussions. This work was supported by NASA Award 80NSSC24K1241.
\end{acknowledgments}

\section*{Data Availability Statement}
The data and code that support the findings of this study are openly available in Zenodo at http://doi.org/10.5281/zenodo.16541374, reference number [\onlinecite{huang2025lmfpccode}].

\appendix

\section{\label{sec:icw_absTime}Dependence of Ion Cyclotron Damping Signatures on System Evolution Duration}
As mentioned at the end of Section~\ref{sec:cd_beta}, to verify whether the increasing deformation of the distribution function and the expansion of FPCs to higher $v_\perp$  in \figref{fig:cd_beta} with increasing $\beta_i$ is driven by the absolute system evolution time, we conduct two additional tests.  By adjusting the  system evolution time $t \Omega_i$ before the FPC correlation interval begins and the length of the FPC 
correlation interval $\tau \Omega_i$, we can compare these new cases to previously calculated cases  with similar values of $t \Omega_i$ and  $\tau \Omega_i$.  We present these values for two new comparisons in Table~\ref{tab:comp_time}.

\begin{figure*}
 \begin{center}
\includegraphics[width=.9\textwidth]{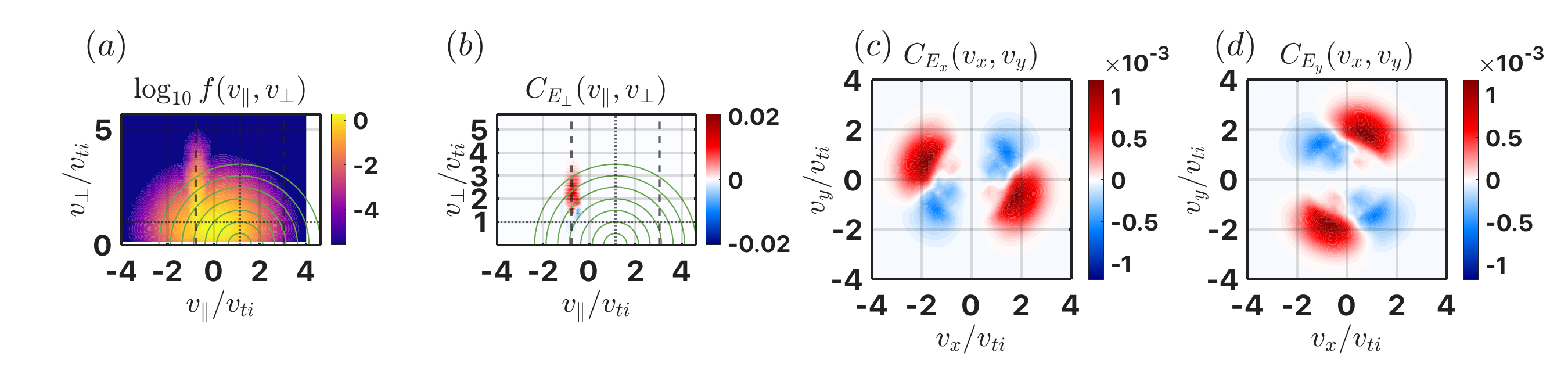}
\includegraphics[width=.9\textwidth]{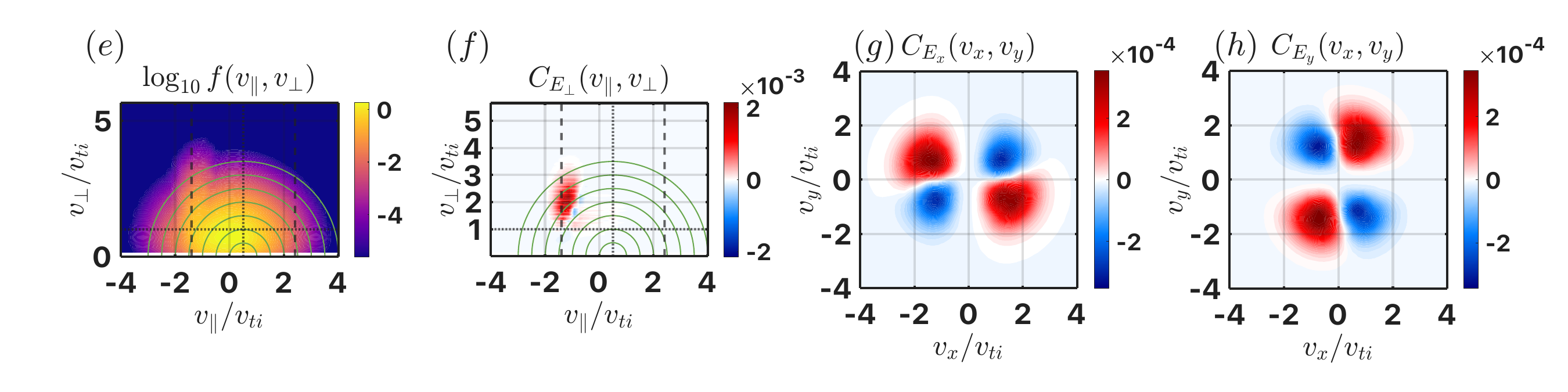}
   \end{center}
 \caption{\label{fig:cd_AbsTimeComparison_b01_b1} Comparison of the velocity-space signatures of ion cyclotron damping between the  $\beta_i = 0.1$ and  $\beta_i = 1$ cases when the  system evolution time $t \Omega_i$ and correlation interval $\tau \Omega_i$ have similar absolute times, as presented in the first two rows of  Table~\ref{tab:comp_time}.}
\end{figure*}

In \figref{fig:cd_AbsTimeComparison_b01_b1}, we present a comparison
between the new $\beta_i=0.1$ case in the first row of
Table~\ref{tab:comp_time} and the old $\beta_i=1$ case from the top row of
\figref{fig:cd_1icw_2icw}.  With the longer absolute
integration time and correlation interval, the gyrotropic signature in
\figref{fig:cd_AbsTimeComparison_b01_b1}(b) for $\beta_i=0.1$ looks
much more qualitatively and quantitatively similar to the $\beta_i = 1$
case in (f).  The perpendicular signatures for  $\beta_i=0.1$ in (c, d) also look more quantitatively similar to the $\beta_i=1$ case in (g, h).
Note that, as emphasized in Section~\ref{sec:lm_misc}, the term "quantitatively similar" here refers specifically to the locations and shapes of the velocity-space patterns, rather than to their absolute amplitudes.

\begin{figure*}
 \begin{center}
\includegraphics[width=.9\textwidth]{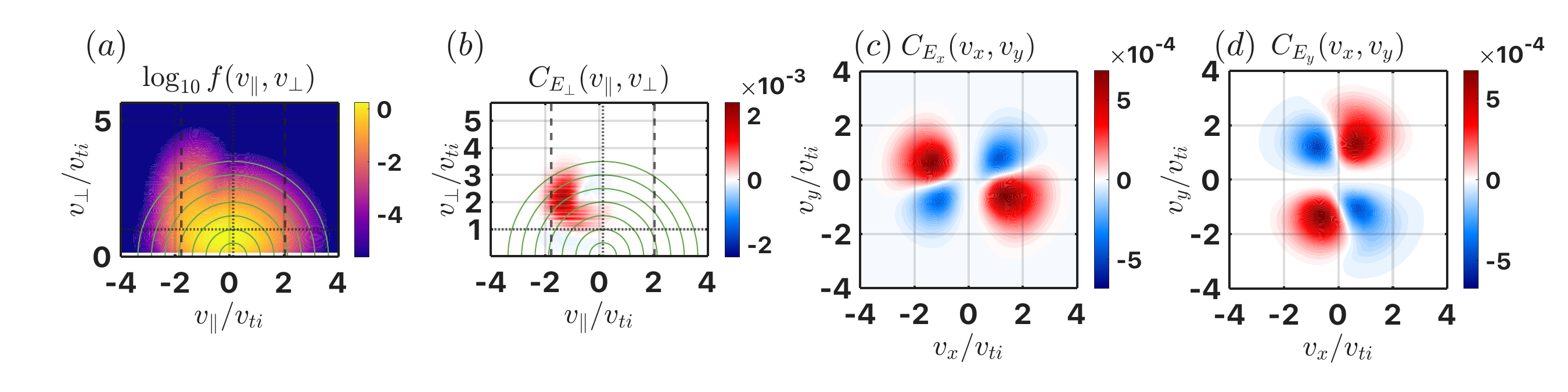}
\includegraphics[width=.9\textwidth]{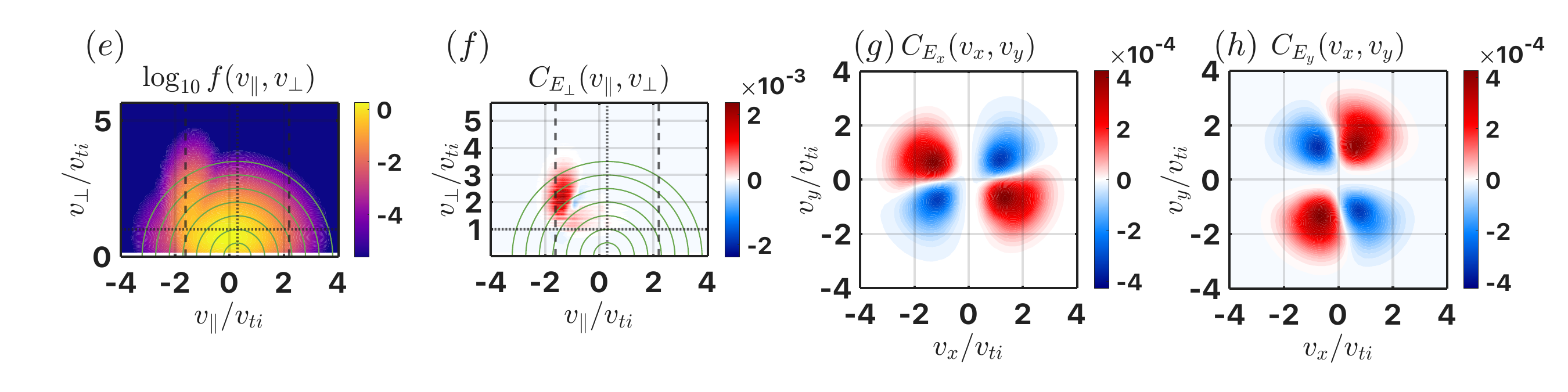}
   \end{center}
 \caption{\label{fig:cd_AbsTimeComparison_b10_b3} Comparison of the velocity-space signatures of ion cyclotron damping between the  $\beta_i = 10$ and  $\beta_i = 3$ cases when the  system evolution time $t \Omega_i$ and correlation interval $\tau \Omega_i$ have similar absolute times, as presented in the last two rows of  Table~\ref{tab:comp_time}.}
\end{figure*}

In \figref{fig:cd_AbsTimeComparison_b10_b3}, we present a comparison
between the new $\beta_i=10$ case in the third row of
Table~\ref{tab:comp_time} and the old $\beta_i=3$ case from the fouth row of
 \figref{fig:cd_beta}.  With the shorter absolute
integration time and correlation interval, the gyrotropic signature in
 \figref{fig:cd_AbsTimeComparison_b10_b3}(b) and the perpendicular signatures in (c, d) for $\beta_i=10$ no longer extend to the much higher $v_\perp$ values observed in  \figref{fig:cd_beta}(r--t), looking much more quantitatively similar to the  $\beta_i=3$ case in  \figref{fig:cd_AbsTimeComparison_b10_b3}(f--h).

Together the results from Figs.~\ref{fig:cd_AbsTimeComparison_b01_b1} and~
\ref{fig:cd_AbsTimeComparison_b10_b3} support our hypothesis that  the
most striking differences, particularly  in the ion cyclotron damping gyrotropic signatures, disappear when their absolute Liouville mapping calculation times are
matched with those of the previous $\beta_i = 1$ and $3$ cases.  This analysis reinforces the prediction from Section~\ref{sec:cd_beta} that  the velocity-space signatures of ion cyclotron damping do not vary significantly with variations in $\beta_i$.

\begin{table}
\begin{ruledtabular}
\begin{tabular}{ccccc}
  $\beta_i$ & $t_\text{init}$ & Interval & $t \Omega_i$ &  $\tau \Omega_i$\\
  \hline
  $0.1$  & $-9T$ & $[0,5T]$ & $95.517$ & $53.065$ \\
  $1$    & $-4T$ & $[0,2T]$ & $94.300$   & $47.150$  \\
  \hline
  $10$   & $-2T$ & $[0,T]$  & $181.324$ & $90.662$ \\
  $3$    & $-4T$ & $[0,2T]$ & $166.688$ & $83.344$ \\
  \end{tabular}
\end{ruledtabular}
\caption{\label{tab:comp_time} Tests to compare cases with different $\beta_i$ (and thus different wave periods) but similar absolute system evolution time before correlation $t \Omega_i$ and correlation interval  $\tau \Omega_i$. The new $\beta_i = 0.1$ case (first row of this table) compares to the $\beta_i = 1$ case from  \figref{fig:cd_1icw_2icw}(a--d), with the comparison shown in \figref{fig:cd_AbsTimeComparison_b01_b1}. The new $\beta_i = 10$ case (third row of this table) compares to the $\beta_i = 3$ case from  \figref{fig:cd_beta}(m--p), with the comparison shown in  \figref{fig:cd_AbsTimeComparison_b10_b3}.}
\end{table}

\section{\label{sec:icd_asymmetry}Asymmetry in the Gyrotropic Velocity-Space Signature of Ion Cyclotron Damping with Two Counter-propagating ICWs}
Revisiting ion cyclotron damping signatures in \figref{fig:cd_1icw_2icw}, it is tempting to predict that, in the case of the standing ICW fields constructed by superposing two counter-propagating ICWs, the resulting gyrotropic FPC signature would simply be the linear superposition of the individual signatures from each single mode.
As shown in \figref{fig:cd_1icw_2icw}, a single propagating ICW produces a signature on one side (with either $v_\parallel >0$ or $v_\parallel <0$) of the gyrotropic velocity plane (near the $n = 1$ resonant velocity of this ICW), while the counterpropagating ICW case yields patterns on both sides, corresponding to the $n = 1$ resonant velocities of the two ICWs. Specifically, the signatures with $v_\parallel >0$ and $v_\parallel <0$ each consist of a red region of increasing phase-space energy density at $v_\perp > v_{ti}$ and a more faint blue region of decreasing phase-space energy density at $v_\perp < v_{ti}$.

\begin{figure*}
 \begin{center}
\includegraphics[width=.9\textwidth]{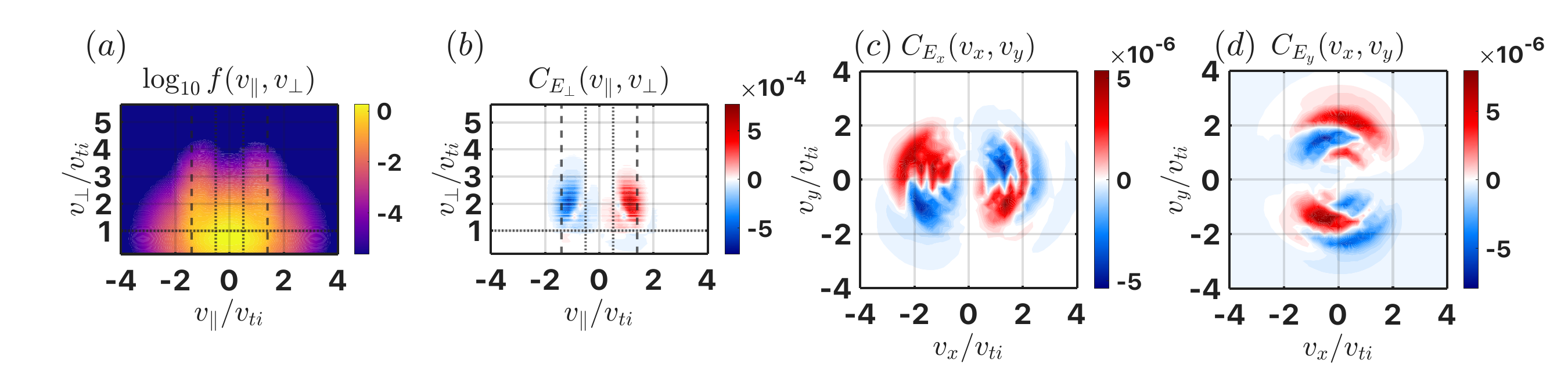}
\includegraphics[width=.9\textwidth]{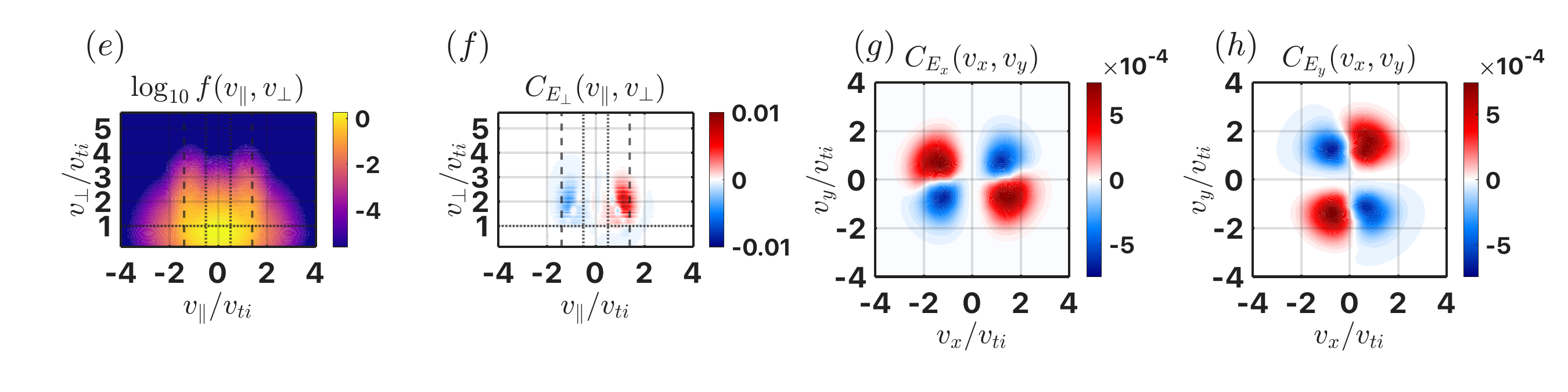}
\includegraphics[width=.9\textwidth]{iSHCDV21_20241118_131107_LF2_ijkl.pdf}
\includegraphics[width=.9\textwidth]{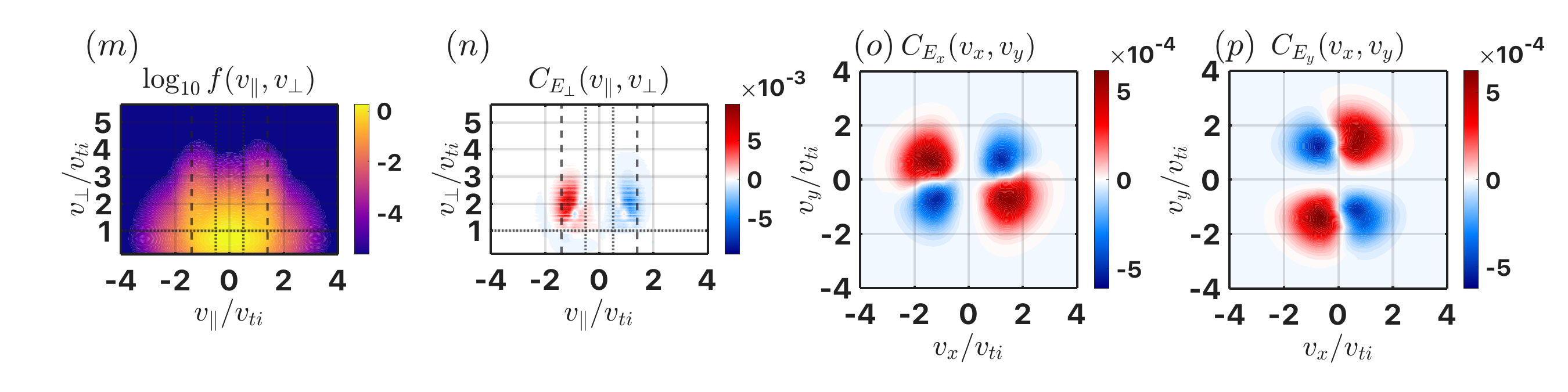}
\includegraphics[width=.9\textwidth]{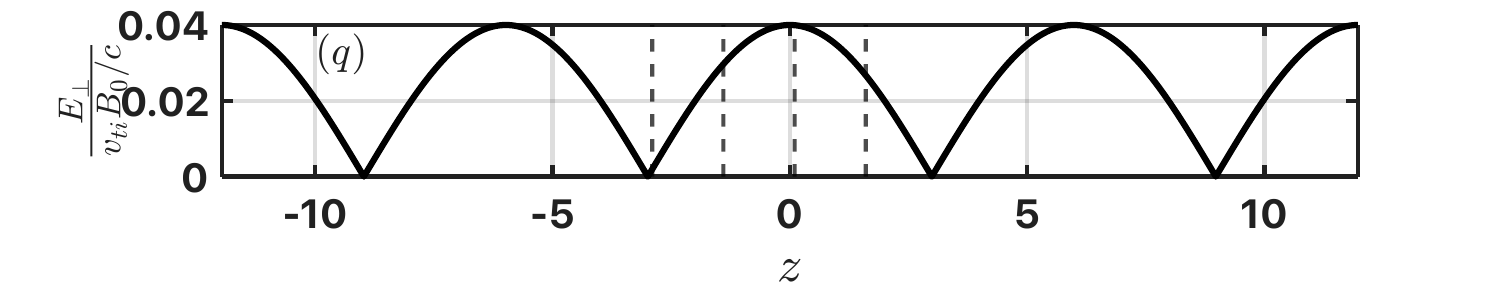}
   \end{center}
\caption{\label{fig:cd_asymmetry} Asymmetrical velocity-space signatures of ion cyclotron damping in a standing wave pattern due to two counter-propagating ICWs, where the FPC analysis (in the same format as \figref{fig:cd_1icw_2icw}) has been performed at $z/\rho_i = -2.9$ (a-d), $-1.4$ (e-h), $0.1$ (i-l), and $1.6$ (m-p).}
\end{figure*}

However, as it turns out, the symmetry of the gyrotropic velocity-space signature for counter-propagating ICWs 
depends on the position of the FPC analysis within the standing, counter-propagating ICW fields, as shown in \figref{fig:cd_asymmetry}. As discussed in Section~\ref{sec:lm_misc}, even though the wave electromagnetic fields are specified from the linear dispersion relation, the Liouville mapping procedure retains nonlinear effects by evolving the distribution function along particle trajectories without dropping nonlinear terms. Thus, the results may not necessarily be interpreted as the sum of purely linear responses. To investigate this further, consider the perpendicular electric field components $E_x(\mathbf{r}, t)$ and $E_y(\mathbf{r}, t)$ generated by two counter-propagating ICWs with wave vectors $\mathbf{k}_{1, 2} \rho_i = 0.01 \hat x \mp 0.525 \hat z$. From the \texttt{PLUME} solver, the two wave modes have equal frequencies $\omega(\mathbf{k}_1) = \omega(\mathbf{k}_2) \equiv \omega$ and Fourier coefficients for the perpendicular components of the electric field $\delta \hat{E}_{x, y} (\mathbf{k}_1) = \delta \hat{E}_{x, y} (\mathbf{k}_2) \equiv \delta \hat{E}_{x, y}$. Writing the Fourier coefficients in polar form, $\delta \hat{E}_{x, y} = |\delta \hat{E}_{x, y}| e^{i \phi_{E_{x, y}}}$, taking the same wave amplitudes with  $\epsilon_1 = \epsilon_2 \equiv \epsilon$,
and substituting into Eq.~(\ref{eq:constructed_em_fields}) and Eq.~(\ref{eq:em_ift}), we obtain the $x$- and $y$-components of the electric field after the first wave period, i.e. once the window function has fully ramped up the amplitude, as follows
\begin{eqnarray}
E_{x,y} (\mathbf{r}, t) = 2 \epsilon |\delta \hat{E}_{x, y}|  \cos\left[\frac{(\mathbf{k}_1 + \mathbf{k}_2) \cdot \mathbf{r} - 2 \omega t + 2\phi_{E_{x, y}}}{2} \right] \nonumber \\
\times \cos \left[\frac{(\mathbf{k}_1 - \mathbf{k}_2) \cdot \mathbf{r}}{2} \right].
\end{eqnarray} 
For a left-hand polarized ICW, $|\delta \hat{E}_x| = |\delta \hat{E}_y|$ and $\phi_{E_x} - \phi_{E_y} = \pi/2$. This gives a time-independent expression for the perpendicular electric field magnitude
\begin{eqnarray}
    E_\perp (\mathbf{r}) = \sqrt{E_x^2(\mathbf{r}, t) + E_y^2 (\mathbf{r}, t)} \nonumber \\
    = 2 \epsilon |\delta \hat{E}_x| \sqrt{\cos^2 \left[\frac{(\mathbf{k}_1 - \mathbf{k}_2) \cdot \mathbf{r}}{2} \right]},
\end{eqnarray}
which varies spatially along $z$ with a wavelength  $\lambda_z=5.984 \rho_i$. As shown in \figref{fig:cd_asymmetry}(q), $E_\perp$ peaks at $z = 5.984 n \rho_i$, where $n = 0, \pm 1, \pm 2, \ldots$, and vanishes at integer multiples of $z = \pm 2.992 \rho_i$.
To examine how this spatial variation affects the FPC signature, we run Liouville mapping at four evenly spaced $z$ positions, $z/\rho_i = -2.9, -1.4, 0.1, 1.6$, while keeping all other parameters identical to those used in the main text. These positions, marked with black dashed lines in \figref{fig:cd_asymmetry}(q), are chosen to avoid the exact node where $E_\perp = 0$. The resulting gyrotropic FPCs are shown in the first through 
fourth rows of \figref{fig:cd_asymmetry}.

At $z = 0.1 \rho_i$ in \figref{fig:cd_asymmetry}(i-l), near the peak $E_\perp$, the FPC shows a symmetric pattern of ion energization at both $v_\parallel >0$ and $v_\parallel <0$. These patterns are centered near the two $n = 1$ resonant velocities associated with the two counter-propagating wave modes. This symmetric feature indicates that ions resonating with both wave modes are gaining energy.

However, at $z = -2.9 \rho_i$ in \figref{fig:cd_asymmetry}(a-d) and $-1.4 \rho_i$ in \figref{fig:cd_asymmetry}(e-h), the FPC pattern becomes asymmetric: ions near the $n = 1$ resonance of the backward-propagating wave ($\mathbf{k}_1$) gain energy, while those near the forward-propagating wave ($\mathbf{k}_2$) lose energy. At $z = 1.6 \rho_i$ in \figref{fig:cd_asymmetry}(m-p), this asymmetry is reversed in $v_\parallel$, with ions near the $\mathbf{k}_2$ resonance gaining energy while those near the $\mathbf{k}_1$ resonance lose energy. The origin of this asymmetry is unclear, and will require further investigation. We suspect that the asymmetry results from nonlinear effects captured by the Liouville mapping method.

\nocite{*}
\bibliography{lmfpcref}

\end{document}